\documentclass{gGAF2e}
\usepackage{lineno}
\usepackage{xcolor}
\usepackage{graphicx}
\providecommand\n{\boldsymbol{\nabla}}
\providecommand\bcdot{{\bm \cdot}}

\newcommand\ncurl{{\bm \nabla\times}}
\newcommand\ndiv{{\bm {\nabla\cdot}}}
\newcommand\rO{{\mathrm O}}
\newcommand\rd{{\mathrm d}}
\newcommand\ri{{\mathrm i}}
\newcommand\K{{\mathrm K}}
\newcommand\I{{\mathrm I}}

\renewcommand{\vec}[1]{\mbox{\boldmath $#1$}}

\def \n {\vec{\nabla}\!}

\def \Om {{\it \Omega}}
\def \rin {r_{\rm in}}
\def \mur {r_{\rm in}}
\def \Rey {\ensuremath{\rm{Re}}}
\def \Ha {\ensuremath{\rm{Ha}}}
\def \Hamin {\ensuremath{\rm{Ha_{min}}}}
\def \Smin {\ensuremath{\rm{S_{min}}}}
\def\HaMIN{\rm Ha_{MIN}}
\def\SMIN{\rm S_{MIN}}
\def \Pm {\ensuremath{\rm{Pm}}}
\def \Rm {\ensuremath{\rm{Rm}}}

\def \S {\ensuremath{\rm{S}}}
\def \Lu {\ensuremath{\rm{S}}}

\def\hats{\hat\sigma}
\def\beg{\begin{equation}}
\def\ende{\end{equation}}
\newcommand{\gsim}{\lower.7ex\hbox{$\;\stackrel{\textstyle>}{\sim}\;$}}
\newcommand{\lsim}{\lower.7ex\hbox{$\;\stackrel{\textstyle<}{\sim}\;$}}

\def\Omin{{\it \Omega}_{\rm in}}
\def\Omout{{\it \Omega}_{\rm out}}

\def\Rout{R_{\rm out}}
\def\Rin{R_{\rm in}}

\def\i{{\rm i}}

\def\omdr{\omega_{\rm dr}}

\def\ara\&a{ Ann. Rev. Astronomy Astrophysics}

\begin{document}

\jvol{00} \jnum{00} \jyear{2017} 

\markboth{G.R\"udiger et al.}{Geophysical and Astrophysical Fluid Dynamics}


\title{Magnetorotational instability
in Taylor-Couette flows between cylinders with finite electrical conductivity
$\!\!^\dag$\footnote{$^\dag$Dedicated to the 90$^{\rm th}$ birthday of P.H.~Roberts}}

\author{G. R\"udiger$^1$$^\ast$ \thanks{$^\ast$Corresponding author. Email: gruediger@aip.de
\vspace{6pt}}, M. Schultz$^1$, F. Stefani$^2$, R. Hollerbach$^3$\\\vspace{6pt} 
$^1$Leibniz-Institut f\"ur Astrophysik Potsdam, An der Sternwarte 16, D-14467 Potsdam, Germany\\
$^2$Helmholtz-Zentrum Dresden-Rossendorf, Bautzner Landstrasse 400, 01328 Dresden, Germany\\
$^3$Department of Applied Mathematics, University of Leeds, Leeds, LS2 9JT, UK\\
 \vspace{6pt}
\received{January 2018} }

\maketitle

\begin{abstract}
The nonaxisymmetric azimuthal magnetorotational instability is studied for hydromagnetic Taylor-Couette flows between cylinders of finite electrical conductivity. We find that the magnetic Prandtl number $\Pm$ determines whether perfectly conducting or insulating boundary conditions lead to lower Hartmann numbers for the onset of instability. Regardless of the imposed rotation profile, for small $\Pm$ the solutions for perfectly conducting cylinders become unstable for weaker magnetic fields than the solutions for insulating cylinders. The critical Hartmann and Reynolds numbers form monotonic functions of the ratio $\hats$ of the electrical conductivities of the cylinders and the fluid, such that $\hats=\rO(10)$ provides a very good approximation to perfectly conducting cylinders, and $\hats=\rO(0.1)$ a very good approximation to insulating cylinders. These results are of particular relevance for the super-rotating case where the outer cylinder rotates faster than the inner one; in this case the critical onset values are substantially different for perfectly conducting versus insulating boundary conditions. An experimental realization of the super-rotating instability, with liquid sodium as the fluid and cylinders made of copper, would need an electric current of at least 33.5~kA running along the central axis.
\end{abstract}
\section{Introduction}
Instabilities in rotating conducting fluids under the influence of magnetic fields have recently attracted considerable interest. In view of astrophysical applications, the consideration of differential rotation is relevant. When considered separately, an axial field and a Keplerian rotation are each stable. Also the stability against axisymmetric perturbations of differential rotation in the presence of {\em azimuthal} magnetic fields in cylindrical geometry has been studied by several authors starting with \cite{V59} and \cite{C61}. Moreover, for purely azimuthal fields which are current-free in the fluid { \em all} Taylor-Couette flows with negative shear (${\rm d}\Om/{\rm d}R<0$) are stable against axisymmetric perturbations \citep{HS06}. In a series of papers we have shown that hydromagnetic Taylor-Couette flows under the influence of such azimuthal magnetic fields are generally unstable against {\em nonaxisymmetric} perturbations. This is in particular true for all rotation profiles $\Om(R)$ decreasing with radius (`sub-rotation') but also, under certain circumstances, for rotation profiles increasing with radius (`super-rotation'). Because of the current-free, and thus force-free, character of the applied magnetic fields, we have called them Azimuthal MagnetoRotational Instability (AMRI), which are nonaxisymmetric by definition.

There are several types of AMRI. If the rotation profile is the potential flow $\Om\propto 1/R^2$ (also named the Rayleigh limit,{  where the specific angular momentum is radially uniform}), then the radial profiles $B_\phi(R)$ and $U_\phi=R\Om(R)$ are identical, as the radial profile of a current-free azimuthal field is $B_\phi\propto 1/R$. Such combinations of $\vec U$ and $\vec B$ were first
 considered by Chandrasekhar, who showed them to be stable in the diffusionless limit \citep{C56}. 
 One can show, however, that real flows of Chandrasekhar-type with finite molecular diffusivities are always unstable against nonaxisymmetric perturbations 
 \citep{RS15}. Curves of neutral stability for $m=1$ in the Hartmann number/Reynolds number coordinate system converge for small magnetic Prandtl numbers,
\beg
{\Pm}\, =\,{\nu}/{\eta}\,,
\label{pmorg}
\ende
or in other words, the eigenvalues scale with Hartmann number and Reynolds number for $\Pm\to 0$.

For flatter rotation profiles -- for example the flow with $\Om\propto 1/R$, i.e. uniform linear velocity $U_\phi=R\Om(R)$ -- the scaling for small $\Pm$ changes dramatically. The neutral stability curves now converge in the ($\Lu/\Rm$) plane formed by the Lundquist number and the magnetic Reynolds number as the ccordinates. Note that this is { - except for the Rayleigh limit -} also true for the standard magnetorotational instability (MRI) with axial magnetic fields. { The immediate consequence is that for small $\Pm$ -- the relevant limit for liquid metals -- laboratory experiments to probe the existence of MRI or AMRI are very difficult. Such experiments, however,  are needed to get data for the instability-induced angular momentum transport in astrophysically relevant parameter regimes 
that are hardly accessible to numerics  \citep[see][]{RG18}}.

The combination of current-free azimuthal magnetic fields and  radially increasing $\Om(R)$ can also be unstable, but there are  differences to the AMRI for sub-rotation. Super-AMRI is a double-diffusive instability,  as it does not exist for $\Pm=1$. { For small $\Pm$ it scales with the Reynolds number and the Hartmann number while for $\Pm\to\infty$ the magnetic Reynolds number and the Lundquist number converge. The Reynolds number remains thus  finite for $\Pm\to 0$ what seems to be  the necessary condition that  experiments with liquid metals as the conducting fluid are possible}. 

Moreover, one finds that for small $\Pm$ the minimal Hartmann numbers $\Hamin$ for perfectly conducting cylinders are {\em much} smaller than for insulating cylinders \citep{RS19}. Almost all theoretical or numerical investigations  worked so far with these extremes as the possible boundary conditions, with the material making up the cylinder walls assumed to be either perfectly conducting or insulating. These assumptions, however, are far from realistic. As an example, the conductivity of copper (as the cylinder material) is only about five times higher than that of liquid sodium (as the conducting fluid), hence the conductivity ratio 
\beg
\hat\sigma\,=\,{\sigma_{\rm cyl}}\,\big/{\sigma_{\rm fluid}}
\label{sigma}
\ende
for this combination approaches the value of (say) 5. The question is whether such a finite  conductivity ratio of cylinders and fluid leads to magnetic fields critical for the onset of instability close to the results for perfectly conducting material or not. In the present paper the influence of finite conductivity of the cylinder walls on the stability of Taylor-Couette flows under the influence of azimuthal magnetic fields is thus considered. The  extremes of the parameter (\ref{sigma}) are $\hats=0$ for insulating cylinders and $\hats=\infty$ for perfectly conducting cylinders. A similar problem for the nonaxisymmetric  instability for an axisymmetric dynamo model using finite   conductivity ratios (\ref{sigma})  has been formulated by  \cite{GF80}.

The equations of the problem are the MHD equations
\begin{subequations}
 \label{mhd2}  
\begin{align}
 \frac{\partial \vec{U}}{\partial t} + (\vec{U}\bcdot \n)\vec{U} \,=\,&\,-\frac{1}{\rho} \n P + \nu \Delta \vec{U} 
 + \frac{1}{\mu_0\rho}{({\ncurl}}\vec{B}) \times \vec{B}\,,\\
 \frac{\partial \vec{B}}{\partial t}\,=\,&\, {{\ncurl}} (\vec{U} \times \vec{B}) + \eta \Delta\vec{B} 
\end{align}
\end{subequations}
along with $ {{\ndiv}}\ \vec{U} = {{\ndiv}}\ \vec{B} = 0$. Here $\vec{U}$ is the velocity of the conducting fluid, $\vec{B}$ the magnetic field, $P$ the pressure, $\rho$ the density, $\nu$ its kinematic viscosity and $\eta$ the magnetic diffusivity (which is inversely proportional to the electric  conductivity $\sigma$). The basic state in cylindrical coordinates ($R,\phi,z$) is \mbox{$ U_R=U_z=B_R=B_z=0$} for the poloidal components and
\beg
\Om \,=\, a_\Om + \frac{{b_\Om}}{R^2}
\label{OM}
\ende
for the rotation profile, with the constants 
\begin{eqnarray}
 a_\Om\,=\,\frac{\mu-\mur^2}{1-\mur^2}\Om_{\rm in}\,, \hskip 20mm
 b_\Om\,=\, \frac{1-\mu}{1-\mur^2}\Om_{\rm in} R_{\rm in}^2\,.
 \label{ab} 
\end{eqnarray}
Here $\mur={R_{\rm in}}/{R_{\rm out}}$ is the ratio of the two cylinder radii. The gap width between the cylinders is $D=\Rout-\Rin$. $\Om_{\rm in}$ and $\Om_{\rm out}$ are the angular velocities of the inner and outer cylinders, respectively. Defining the ratio $\mu= \Om_{\rm out}/\Om_{\rm in}$, sub-rotation is represented by $\mu<1$ and super-rotation by $\mu>1$.

For the magnetic field the stationary solution is 
\begin{eqnarray}
 B_\phi\,=\,a_B R+\frac{b_B}{R}\,.
 \label{basicB} 
\end{eqnarray}
Only the field with the radial profile $B_\phi\propto 1/R$ is current-free in the fluid. We define $\mu_B=B_{\rm out}/B_{\rm in}$. The current-free field is then given by $\mu_B=\rin$. In most cases we take $\rin=0.5$, but occasionally narrow gaps with $\rin=0.9$ are also considered.

The dimensionless physical parameters of the system are the 
Hartmann number $\Ha$ and the Reynolds number $\Rey$, i.e. 
\begin{eqnarray}
 {\Ha}\, =\,\frac{B_{\rm in} R_0}{\sqrt{\mu_0\rho\nu\eta}}\,, \hskip 20mm
 {\Rey}\, =\,\frac{\Om_{\rm in} R_0^2}{\nu},
\label{pm}
\end{eqnarray}
together with the magnetic Prandtl number (\ref{pmorg}). The Hartmann number is defined at the inner boundary, where the field is strongest. The associated definitions $\Rm=\Pm\ \Rey$ for the magnetic Reynolds number and $\Lu=\sqrt{\Pm}\ \Ha$ for the Lundquist number also yield important parameters that are independent of the viscosity. The quantity $R_0=\sqrt{\Rin D}$ represents the unit of radial distances.

For the stability analysis the variables $\vec{U}$, $\vec{B}$ and $P$ are split into mean and fluctuating components $\vec{U}=\bar{ \vec{U}}+\vec{u}$, $\vec{B}=\bar{ \vec{B}}+\vec{b}$ and $P=\bar P+p$. The bars from the variables are immediately dropped, so that the upper-case letters $\vec{U}$, $\vec{B}$ and $P$ represent the background quantities. By developing the disturbances $\vec{u}$, $\vec{b}$ and $p$ into normal modes 
\beg
\Big[\vec{u},\vec{b},p\Big]=\Big[\vec{u}(R),\vec{b}(R),p(R)\Big] {\rm e}^{{\rm i}(\omega t+kz+ m\phi)}\,,
\label{fluc}
\ende
the solutions of the linearized MHD equations are considered for axially unbounded cylinders. Here $k$ is the axial wave number, $m$ the azimuthal wave number and $\omega$ the complex frequency, including growth rate as its (negative) imaginary part and a drift frequency $\omega_{\rm dr}$ as its real part. A linear code is used to solve the resulting set of linearized ordinary differential equations for the radial functions of flow, field and pressure fluctuations. The solutions are optimized with respect to the Reynolds number for given Hartmann number by varying the wave number. Only the solutions for $m=1$ are discussed, where the sign of $m$ does not play a role here, since `right' and `left' spirals are identical in a purely azimuthal background field. The hydrodynamic boundary conditions at the cylinder walls are no-slip, $u_R=u_\phi=u_z=0$.
\section{Magnetic boundary conditions}
We have a fluid with magnetic diffusivity $\eta_{\rm fluid}$ contained within $R_{\rm in}\le R\le R_{\rm out}$. The regions $R<R_{\rm in}$ and $R>R_{\rm out}$ are solid material with magnetic diffusivity $\eta_{\rm cyl}$, and undergoing solid-body rotation at $\Omin$ for the inner region and $\Omout$ for the outer region. Following  \cite{R64}, we then wish to derive the appropriate boundary conditions to be applied on the magnetic field at $R_{\rm in}$ and $R_{\rm out}$.

We start with the exterior region $R>R_{\rm out}$. The induction equation for the field in this region is
\begin{equation}
\frac{\partial{\vec b}}{\partial t}={{\ncurl}}({\vec U}\times {\vec b})+\eta_{\rm cyl}
\Delta{\vec b},
\label{equa1}
\end{equation}
with ${\vec U}=\Omout R\,{\bf \hat e}_\phi$. Using also the modal expansion (\ref{fluc}), this becomes
\begin{equation}
\ri\hat\omega{\vec b}\,=\,\Delta{\vec b}\,,\hskip 15mm{\rm where} \hskip 15mm
 \hat\omega=\frac{\omega+m\Omout}{\eta_{\rm cyl}}\,.
\label{equa2}
\end{equation}
The physical interpretation of the $m\Omout$ term is that it represents the solid-body rotation of the phase in the moving conducting region.

The three components of (\ref{equa2}) can be expressed as
\begin{align}
\ri\hat\omega b_\pm\, = \,&\,\biggl(\frac{\rd^2}{\rd R^2}+\frac{1}{R}\frac{\rd}{\rd R}\biggr)
b_\pm - \biggl(k^2 + \frac{(m\pm1)^2}{R^2}\biggr)b_\pm\,,  \\
\ri\hat\omega b_z\, = \,&\,\biggl(\frac{\rd^2}{\rd R^2}+\frac{1}{R}\frac{\rd}{\rd R}\biggr)
b_z - \biggl(k^2 + \frac{m^2}{R^2}\biggr)b_z\,,
\end{align}
where $b_\pm=b_R\pm \ri b_\phi$. Solutions decaying at $R\to\infty$ are the modified Bessel functions of the second kind $b_\pm = c_\pm \K_{m\pm1}(\kappa R)$ and $b_z=c_z \K_m(\kappa R)$,
where
\begin{equation}
\kappa\,=\,\sqrt{k^2 + \ri\hat\omega}\,.
\label{kappa1}
\end{equation}

By construction these solutions satisfy (\ref{equa1}). They must also satisfy $ {{\ndiv}}{\vec{b}}=0$. After some algebra this becomes
\begin{eqnarray}
 c_+\bigl(\kappa R \K'_{m+1} + (1+m)\K_{m+1}\bigr)
+c_-\bigl(\kappa R \K'_{m-1} + (1-m)\K_{m-1}\bigr)
+c_z\bigl(2\ri kR\K_m\bigr)=0\,,\hskip 15mm 
\end{eqnarray}
where primes denote the derivatives of the $\K_{m\pm1}$. Using the recursion relations
\begin{equation}
\K'_m(\xi)\,=\,-\,\K_{m\pm1}(\xi) \pm \frac{m}{\xi}\K_m(\xi)
\end{equation}
this simplifies to
$c_z={\kappa}\bigl(c_+ + c_-\bigr)\big/({2\ri k})$.
There are thus two linearly independent solutions, which we can take to be ${\vec B}_1$, defined by $c_+=c_-$, and ${\vec B}_2$, defined by $c_+=-c_-$. Some straightforward algebra, again also using the appropriate recursion relations, then yields
\begin{align}
{\vec B}_1\,=\,&\,\biggl(\K'_m\,,\,\frac{\ri m}{\kappa R}\K_m\,,\,\frac{\ri\kappa}{k}\K_m\biggr),  &
{{\ncurl}}{\vec B}_1\,=\,&\,\biggl(-\,\frac{\ri\hat\omega m}{k\kappa R}\K_m\,,\,
\frac{\hat\omega}{k}\K'_m\,,\,0\biggr),\hskip 2mm\\
{\vec B}_2\,=\,&\,\biggl(-\,\frac{\ri m}{\kappa R}\K_m\,,\,\K'_m\,,\,0\biggr), &
{{\ncurl}}{\vec B}_2\,=\,&\,\biggl(-\ri k\K'_m\,,\,\frac{km}{\kappa R}\K_m\,,\,\kappa \K_m\biggr),\hskip 2mm
\end{align}
where the $\K_m$ are all evaluated at $\kappa R$.

Having derived the solutions in the region $R>R_{\rm out}$, we next consider the matching conditions that must be satisfied at $R=R_{\rm out}$. The usual electromagnetic boundary conditions are that all three components of the magnetic field, as well as the tangential components of the electric field, must be continuous across the interface. From Ohm's law we have that ${\vec E}=\sigma^{-1}{\vec J}-{{\vec U}\times{\vec B}}$. Since all three components of $\vec U$ and $\vec B$ are continuous, continuity of the tangential components of $\vec E$ implies continuity of the tangential components of $\sigma^{-1}{\vec J}\sim\sigma^{-1}{{\ncurl}}{\vec B}$.

If $(b_R,b_\phi,b_z)$ represents the field in the fluid, and $c_1{\vec B}_1 + c_2{\vec B}_2$ represents a general field in $R>R_{\rm out}$, the five continuity conditions then become
\begin{align}
b_R\,=\,c_1\K'_m - c_2\frac{\ri m}{\kappa R}\K_m\,, \hskip 10mm b_\phi\,=\,&\,c_1\frac{\ri m}{\kappa R}\K_m + c_2\K'_m\,, \hskip 10mm b_z\,=\,c_1\frac{\ri\kappa}{k}\K_m\,,\hskip 5mm
\label{cond1}\\[0.1em]
\sigma^{-1}_{\rm fluid}\Bigl(\ri kb_R-\frac{\rd}{\rd R}b_z\Bigr)
=\,&\,\sigma^{-1}_{\rm cyl}\Bigl(c_1\frac{\hat\omega}{k}\K'_m+c_2\frac{km}{\kappa R}
\K_m\Bigr)\,,
\label{cond4}\\[0.2em]
\sigma^{-1}_{\rm fluid}\Bigl(\frac{1}{R}\frac{\rd }{\rd R}(Rb_\phi)-\frac{\i m}{R}b_R
\Bigr)=\,&\,\sigma^{-1}_{\rm cyl}\Bigl(c_2\kappa \K_m\Bigr)\,,
\label{cond5}
\end{align}
all evaluated at $R_{\rm out}$. To reduce these to boundary conditions involving only the internal field $(b_R,b_\phi,b_z)$, it is obviously straightforward to solve (\ref{cond1}c) for $c_1$ and (\ref{cond5}) for $c_2$; inserting these into (\ref{cond1}a,b) then yields the boundary conditions at $R_{\rm out}$ as
\begin{align}
b_R+\frac{\ri k\K'_m(\kappa R)}{\kappa \K_m(\kappa R)}b_z\,=\,&\,
-\frac{\ri m\hat\sigma}{\kappa^2R^2}\biggl(\frac{\rd}{\rd R}(Rb_\phi)-imb_R\biggr),  
\label{bc1outer}\\[0.2em]
b_\phi-\frac{mk}{\kappa^2 R}b_z\,=\,&\,\frac{\hat\sigma \K'_m(\kappa R)}
{\kappa R \K_m(\kappa R)}\biggl(\frac{\rd}{\rd R}(Rb_\phi)-\ri mb_R\biggr),
\label{bc2outer}
\end{align}
where $\hat\sigma$ is defined with (\ref{sigma}). Note also that the matching condition (\ref{cond4}) appears to be missing here; this is in fact automatically satisfied as well then, by virtue of the fact that the field within the fluid also satisfies ${{\ndiv}}{\vec b}=0$, as well as its
induction equation. We see that we thus have the correct number of outer boundary conditions (\ref{bc1outer}) and (\ref{bc2outer}), formulated entirely in terms of the field within the fluid.

Turning next to the boundary conditions at $R_{\rm in}$, the only difference is that now we require the Bessel functions $\I$ rather than $\K$, to ensure regularity as $R\to0$. The recursion formulas for these are slightly different, involving various $\pm$ interchanges. Once these changes are tracked through
though, the equivalents of ${\vec B}_1$ and ${\vec B}_2$ have the same form, except with $\I_m$ instead of $\K_m$. The boundary conditions at $R_{\rm in}$ are therefore
\begin{align}
b_R+\frac{\ri k\I'_m(\kappa R)}{\kappa \I_m(\kappa R)}b_z\,=\,&\,
-\,\frac{\ri m\hat\sigma}{\kappa^2R^2}\biggl(\frac{\rd}{\rd R}(Rb_\phi)-\ri mb_R\biggr),
\label{bc1inner}\\[0.2em]
b_\phi-\frac{mk}{\kappa^2 R}b_z\,=\,&\,\frac{\hat\sigma \I'_m(\kappa R)}
{\kappa R \I_m(\kappa R)}\biggl(\frac{\rd}{\rd R}(Rb_\phi)-\ri mb_R\biggr),
\label{bc2inner}
\end{align}
\vskip -2mm
\begin{equation}
\kappa=\sqrt{k^2 + \ri\hat\omega}\,,
\hskip 15mm {\rm where}  \hskip 15mm 
\hat\omega=\frac{\omega+m\Omin}{\eta_{\rm cyl}}\,.
\end{equation}
(Note how we use the same $\hat\omega$ and $\kappa$ notation for both inner and outer boundaries, but these refer to different values, since $\hat\omega$ involves $\Omin$ versus $\Omout$ at the two locations.)

To summarize, (\ref{bc1outer}), (\ref{bc2outer}) at $R_{\rm out}$ and (\ref{bc1inner}), (\ref{bc2inner}) at $R_{\rm in}$ are the appropriate boundary conditions to impose if the entire regions $R>R_{\rm out}$ and $R<R_{\rm in}$ are made of material having (nondimensional) conductivity $\hat\sigma=\sigma_{\rm cyl}/\sigma_{\rm fluid}$ (which is inversely proportional to the magnetic diffusivity ratio $\hat\eta=\eta_{\rm cyl}/\eta_{\rm fluid}$). In the limiting cases $\hat\sigma\to0$ or $\hat\sigma\to\infty$ these conditions correctly reduce to the more familiar insulating or perfectly conducting (respectively) boundary conditions. Similarly, for axisymmetric $m=0$ modes they simplify somewhat to more familiar forms.

Note also how the eigenvalue $\omega$ enters into these boundary conditions, and in an extremely complicated way: $\hat\omega$ from (\ref{equa2}) appears in $\kappa$ in (\ref{kappa1}) which is part of the argument of the $\K_m$ in the outer boundary conditions (\ref{bc1outer}) and (\ref{bc2outer}), and similarly for the inner boundary conditions. This convoluted dependence on $\omega$ means that the eigenvalue problem is more complicated than the traditional form $\bf Av=\omega Bv$, where the matrices $\bf A$ and $\bf B$ do not involve $\omega$. Instead, it must be formulated as $\bf Cv=0$, where $\omega$ enters into $\bf C$ in a manner more complicated than simply $\bf C=A-\omega B$. The numerical procedure then involves scanning over both $k$ and $\omega$ to obtain the optimal values satisfying $\det{\bf C}=0$. It is precisely to avoid such additional difficulties, and retain the familiar eigenvalue form $\bf Av=\omega Bv$, that \cite{HC07} impose simplified finitely conducting boundary conditions that are valid only if the regions making up the boundaries are assumed to be of arbitrary conductivity, but much narrower than the gap width $D$, in contrast with the results here, where the entire exterior regions are taken to be the finitely conducting material.

It is useful also to further consider the relations (\ref{equa2}) and (\ref{kappa1}) for $\hat\omega$ and $\kappa$ (and their equivalents at the inner boundary). In normalized form these can be written as
\begin{eqnarray}
R_0^2\kappa^2\,=\,R_0^2 k^2 + \i {\Rm} \frac{ \omega+m \Omout}{\Omin}\ \hat\sigma\,, 
\label{az1}
\end{eqnarray}
where again $\Rm=\Pm\ \Rey$. The second term on the right describes the skin effect in electrodynamics. If we assume for the moment that $\omega/\Omin$ does not depend strongly on $\Pm$, then the limit $\Pm\to 0$ hardly influences this term if the instability scales with $\Rm$ for small $\Pm$. If, on the other hand, the instability scales with $\Rey$ for $\Pm\to 0$, then the skin term becomes small for small $\Pm$. We thus expect in this case no great influence of the cylinder's finite conductivity for small $\Pm$.

In this work we then consider the influence of these finitely conducting boundary conditions on the instability of an azimuthal magnetic field due to an axial electric current inside the inner cylinder, $R<\Rin$. The resulting field is current-free, hence $B_\phi\propto 1/R$ so that $\mu_B=0.5$ for $\rin=0.5$. The latter choice leads to $\Rout=2 \Rin$ and to $R_0=\Rout- \Rin$. This field is known to be unstable for rotation profiles with $\mu\geq 0.25$, which would be stable without magnetic fields. We consider two different rotation profiles, $\mu=0.25$ and $\mu=0.5$. The motivation for considering these two profiles is that for small $\Pm$ the two types of instability behave differently. The steep rotation profile $\mu=0.25$ scales with $\Rey$ and $\Ha$ for $\Pm\to 0$, while the flat rotation profile $\mu=0.5$ scales with $\Rm$ and $\S$ for $\Pm\to 0$. The reason for this differing behavior is that the former one belongs to the class of Chandrasekhar flows satisfying the defining condition $\vec{U}\propto \vec B$. We shall discover that the behavior of the instability maps for finite conductivity of the boundaries also differs for these two types of AMRI.

Also for super-rotation with $\mu>1$ two different rotation laws are considered, this time   in a narrow gap. Of particular interest, of course, is the stability/instability behaviour of a Taylor-Couette flow when  a stationary   inner cylinder is assumed.

\section{Quasi-uniform flow}\label{Quasi}
It is clear from (\ref{OM}) that a uniform flow with $\Om\propto 1/R$ is not an exact realization of the background flow. One can only model a quasi-uniform flow with the same value of $U_\phi$ at the two boundaries by use of $\mu=\rin$, so that $a_\Om=\rin\Omin/(1+\rin)$ and $b_\Om=\Rin^2\Omin/(1+\rin)$. 
\begin{figure}[h]
 \centering
 \vbox{
 \hbox{
 \includegraphics[width=0.49\textwidth]{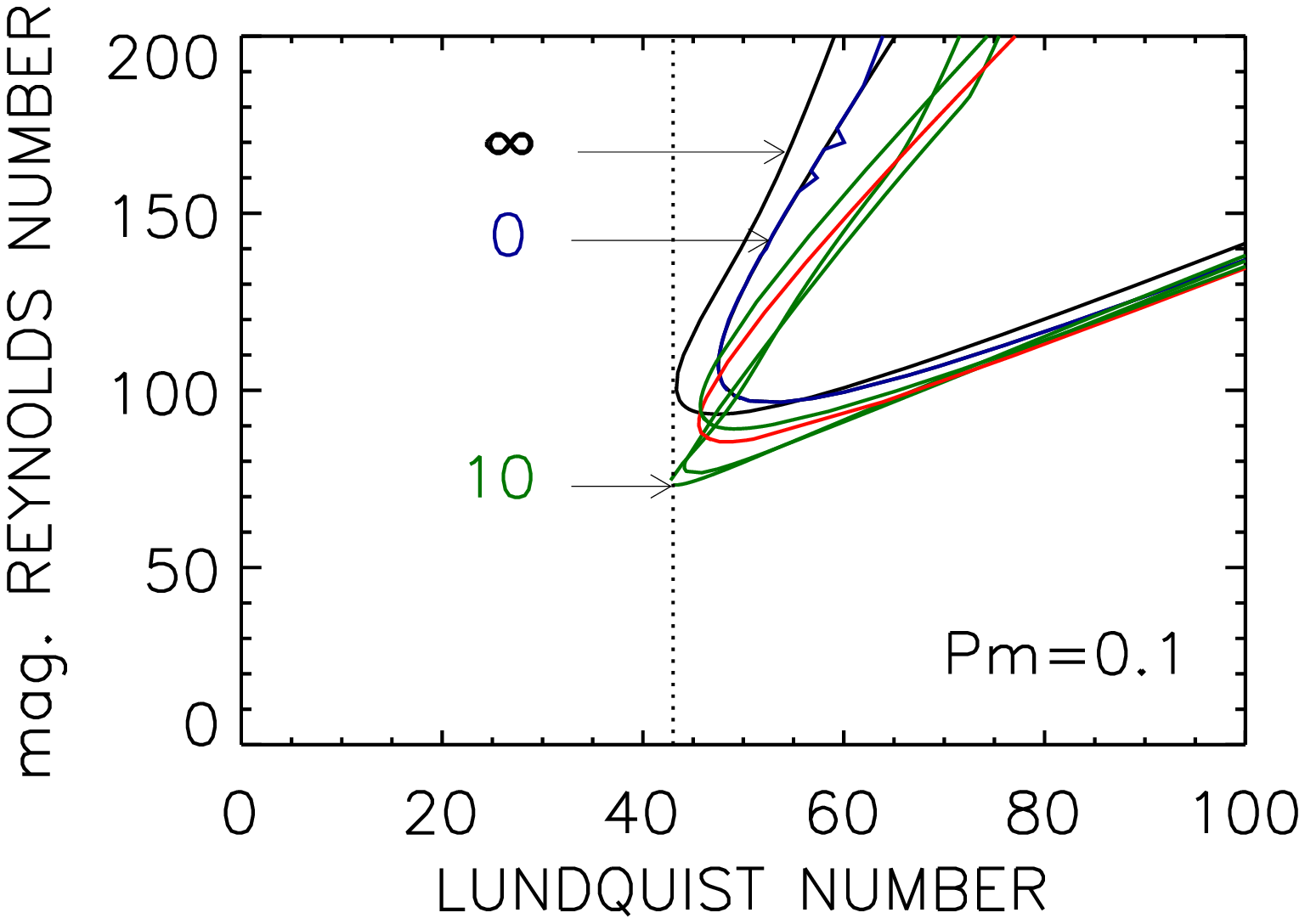}
 \includegraphics[width=0.49\textwidth]{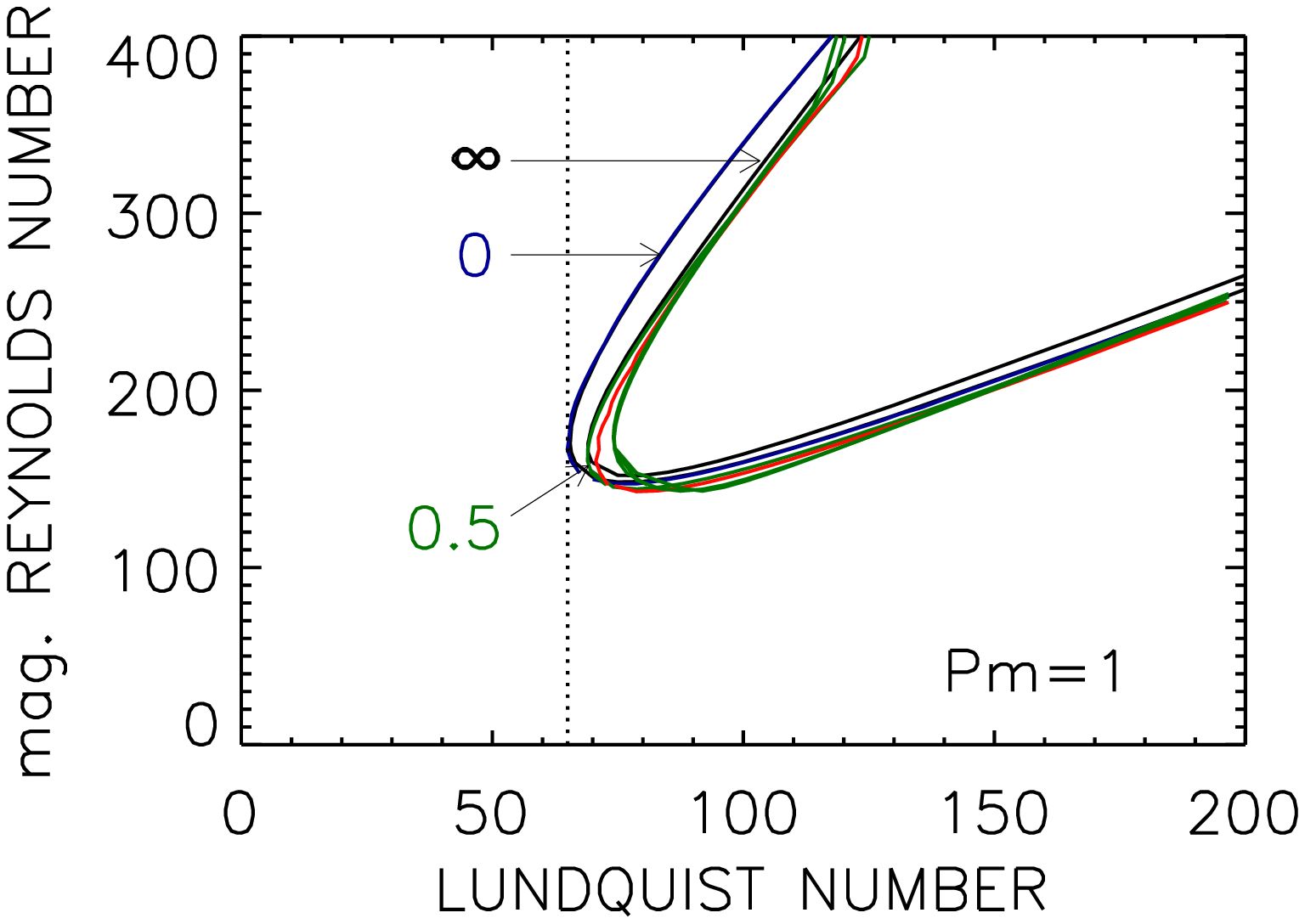}}
 \hbox{\hskip4cm{\includegraphics[width=0.49\textwidth]{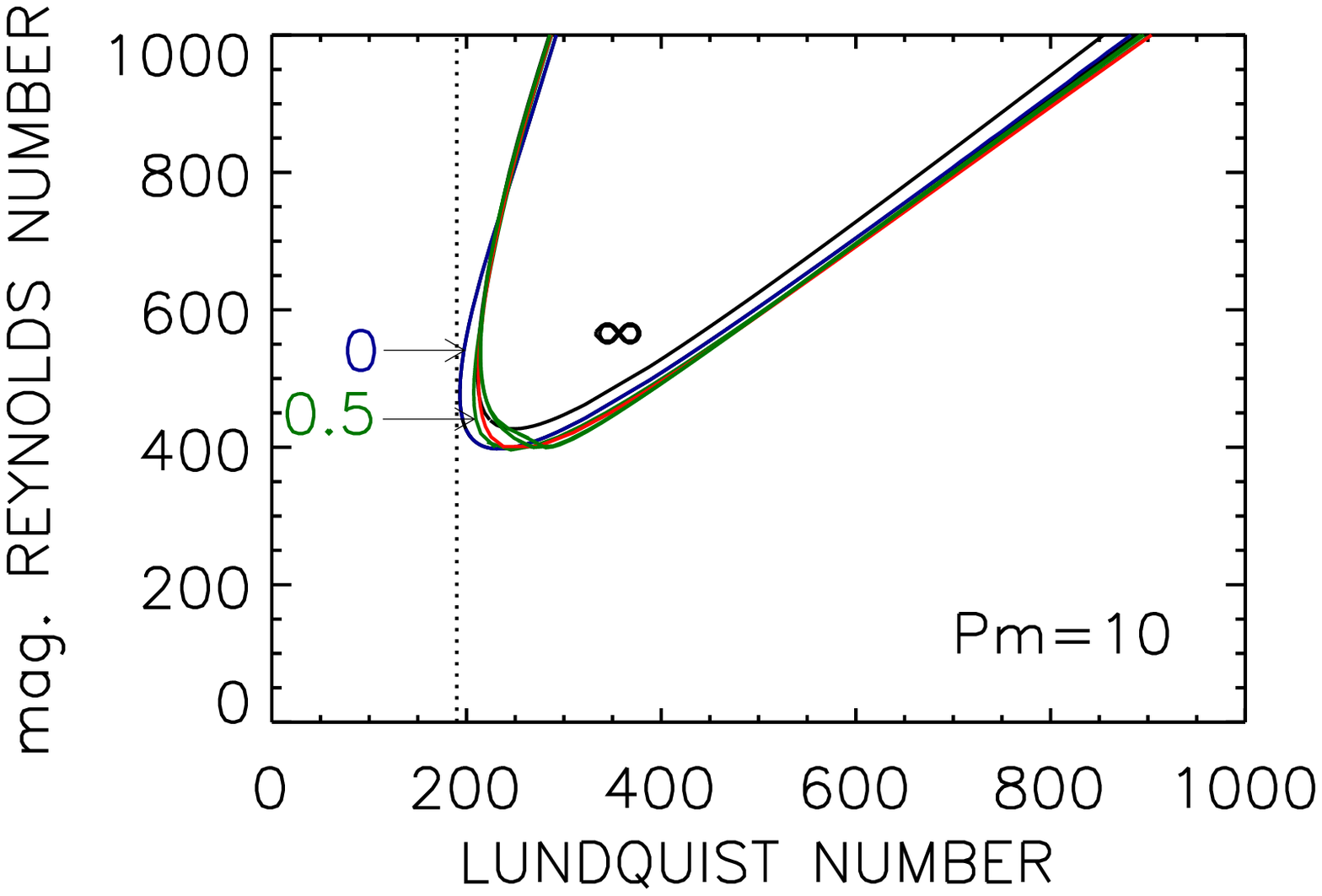}}} 
}
\caption{Stability maps in the ($\S/\Rm$) planes for quasi-uniform rotation for the current-free magnetic field between the cylinders. Top panel:  $\Pm=0.1$, $\Pm=1$, bottom panel: $\Pm=10$. The curves are marked with their values of $\hat\sigma$: $\hat\sigma=0$ (blue), $\hats=0.5$ (green), $\hat\sigma=1$ (red), $\hat\sigma=\infty$ (black). The vertical dotted line indicates the global minimum Hartmann number $\Ha_{\rm MIN}$. $m= 1$. $\mu=\mu_B=\rin=0.5$. (colour online)}
\label{fig1} 
\end{figure}

Figure \ref{fig1} shows the lines of marginal instability for this flow for $\Pm=0.1$  to $\Pm=10$. The curves for insulating cylinders ($\hat\sigma=0$) and for perfectly conducting cylinders ($\hat\sigma=\infty$) are marked by blue and black lines, respectively. They are rather close together, with the insulating material representing the absolutely lowest Lundquist number for $\Pm\geq 1$. For $\Pm<1$ a high conductivity of the cylinder material, with $\hats>10$, makes the system maximally unstable with respect to the absolutely lowest Lundquist number. The corresponding magnetic Reynolds number takes its minimum for $\hats=10$. Higher or lower conductivity ratios lead to higher values of the critical Reynolds number for the onset of instability. These results already suggest the influence of finite conductivity with $\hats>1$ as nontrivial for small $\Pm$.
\begin{figure}[ht]
 \centering
 \hbox{
 \includegraphics[width=0.5\textwidth]{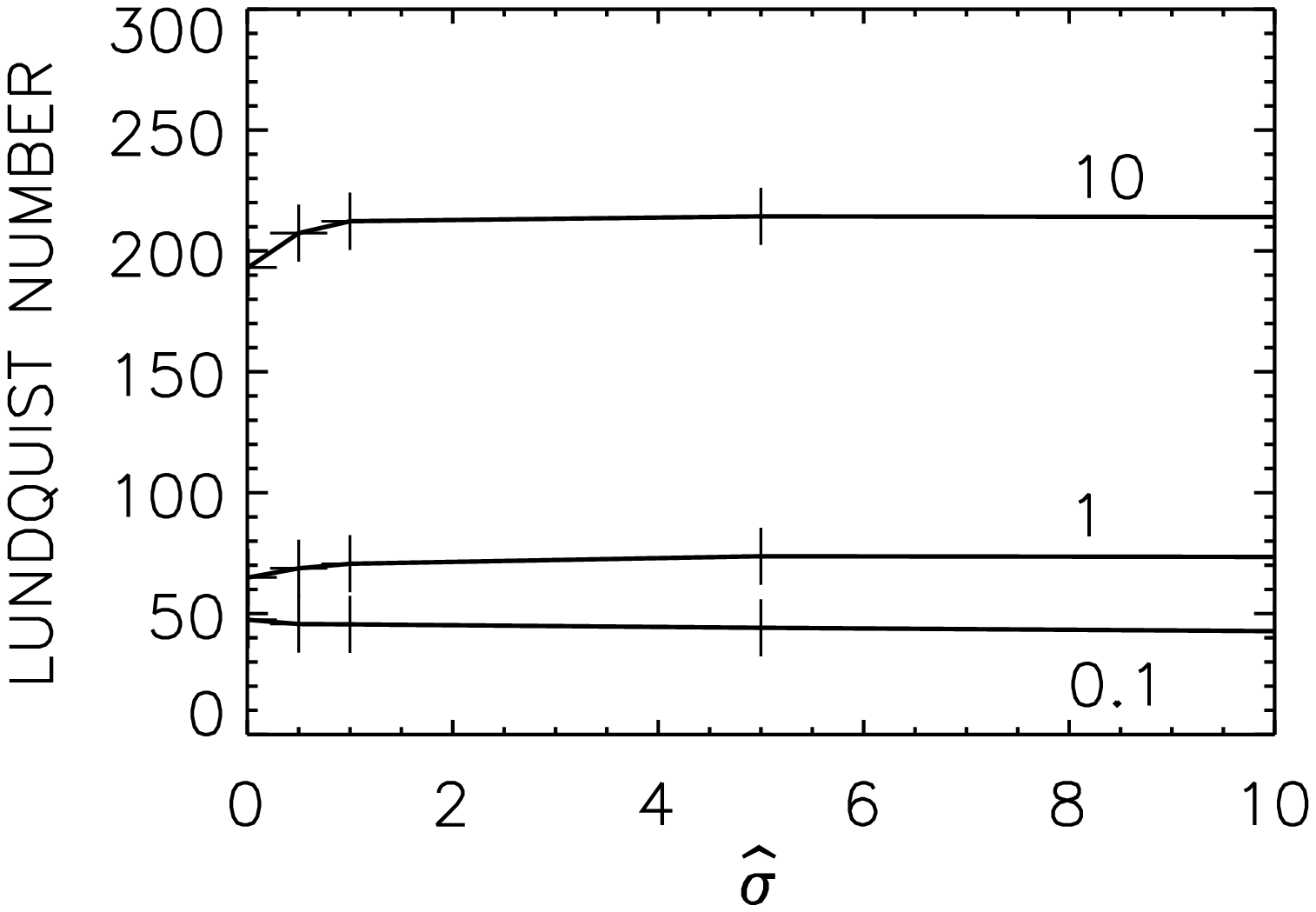} \hfill
 \includegraphics[width=0.5\textwidth]{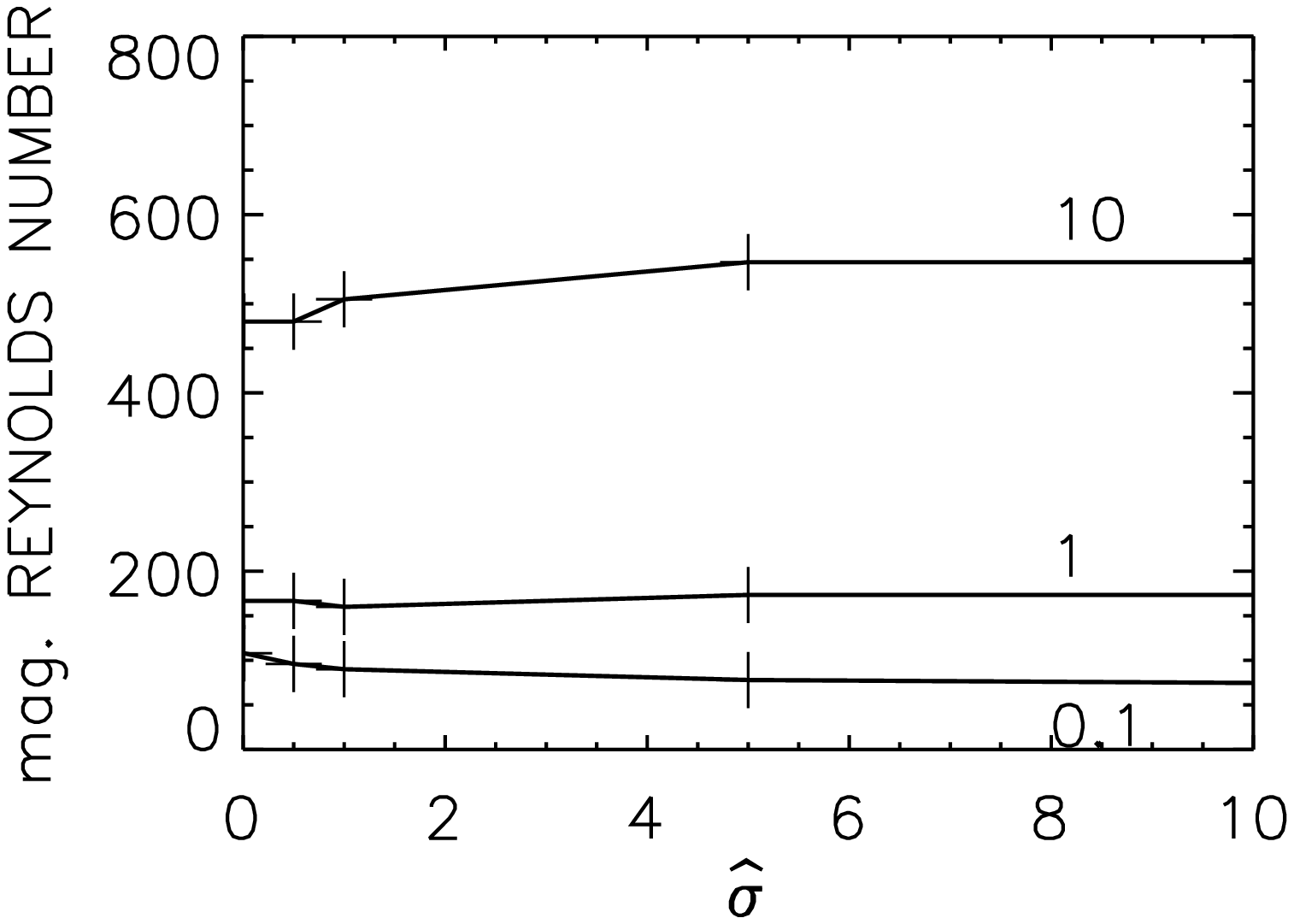} 
 }
 \caption{The coordinates of the minimal magnetic field (left, given as Lundquist number) and the related rotation rate (right, given as magnetic Reynolds number) versus the conductivity ratio $\hat\sigma$ for various magnetic Prandtl numbers ($\Pm= 0.1$, $\Pm=1$  and $\Pm=10$, marked). $\hats=0$ corresponds to insulating cylinders.  $m= 1$, $\mu_B=\mu=\rin=0.5$.}
\label{fig3} 
\end{figure}

Since for $\Pm\to 0$ the lines of marginal stability converge in the ($\Lu/\Rm$) plane, we have given the minimum values in these units. The left panel of figure  \ref{fig3} gives the minimal Lundquist number of the stability lines for various conductivity ratios $\hat\sigma$ and various magnetic Prandtl numbers $\Pm$. We shall focus discussion on those minimal Hartmann or Lundquist numbers above which the MHD flow becomes unstable. To know how this quantity depends on the choice of the magnetic boundary conditions might be important for the experiments. The absolutely smallest of them is called the minimum value $\HaMIN$ or $\SMIN$. We find that these values belong to the perfect-conductor condition for small $\Pm$, and to the insulating condition for large $\Pm$. Note that $\hats=1$ separates two different parts of the function $\Smin=\Smin(\hat\sigma)$ with different slopes, so that the solutions for $\hats\ll 1$ are  very close to the solution for $\hats=0$ (insulating) while those for $\hats\gg 1$ are very close to the solution for $\hats=\infty$ (perfectly conducting). Observe also that at $\hats=0$ the slope of the function $\Lu_{\rm min}(\hats)$ is negative for small $\Pm$. As $\Lu_{\rm min}(\hats)$ is always monotonic it is clear that for small $\Pm$ $\Lu_{\rm MIN}$ always appears for perfectly conducting cylinders. The critical magnetic Reynolds numbers are given in the right panel of figure  \ref{fig3}. The curves are very similar to those for the magnetic field amplitudes (represented by the Lundquist number), for both insulating and conducting cylinders.
\begin{figure}[ht]
 \centering
 \hbox{
 \includegraphics[width=0.49\textwidth]{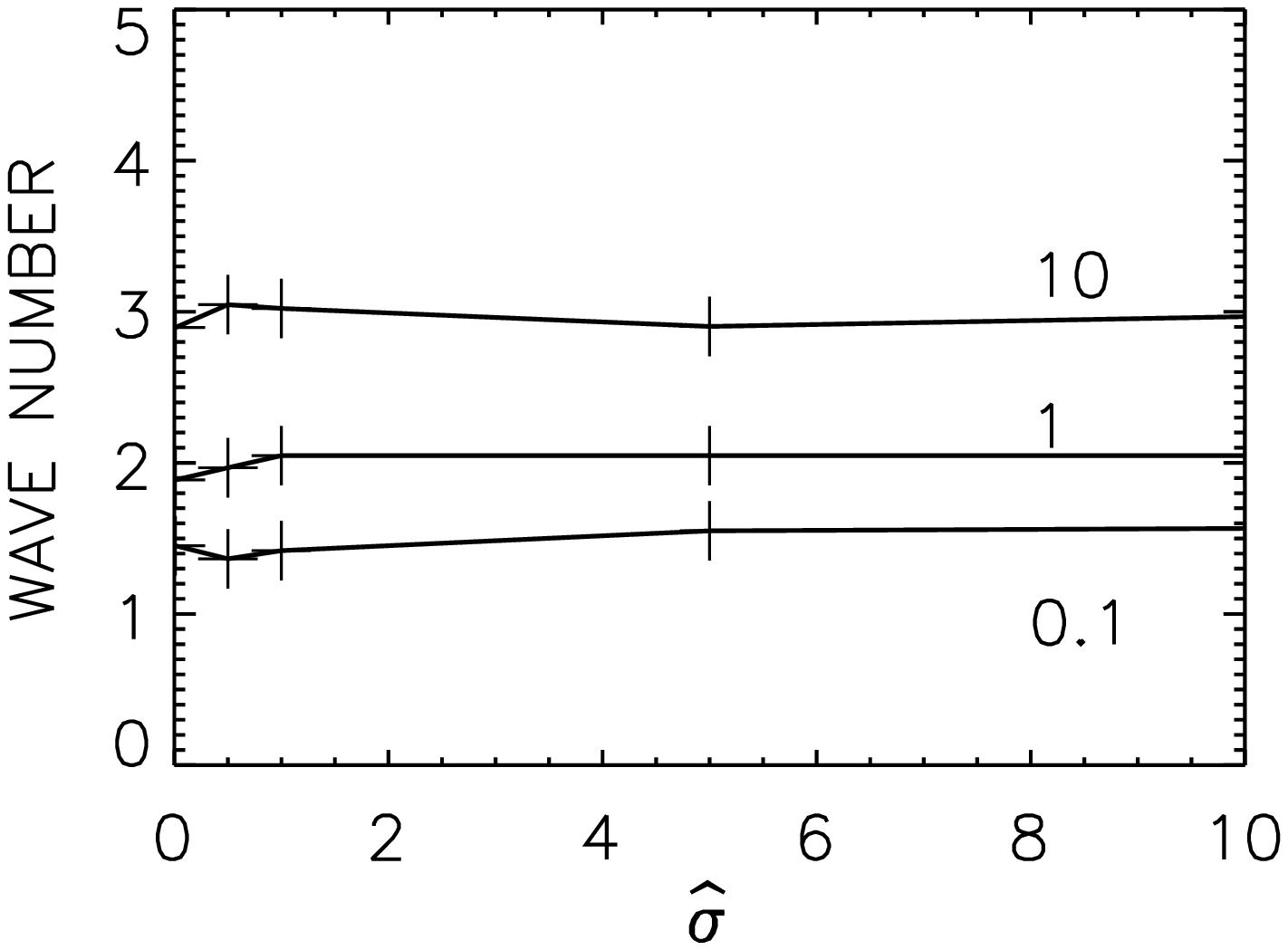}\hfill
 \includegraphics[width=0.49\textwidth]{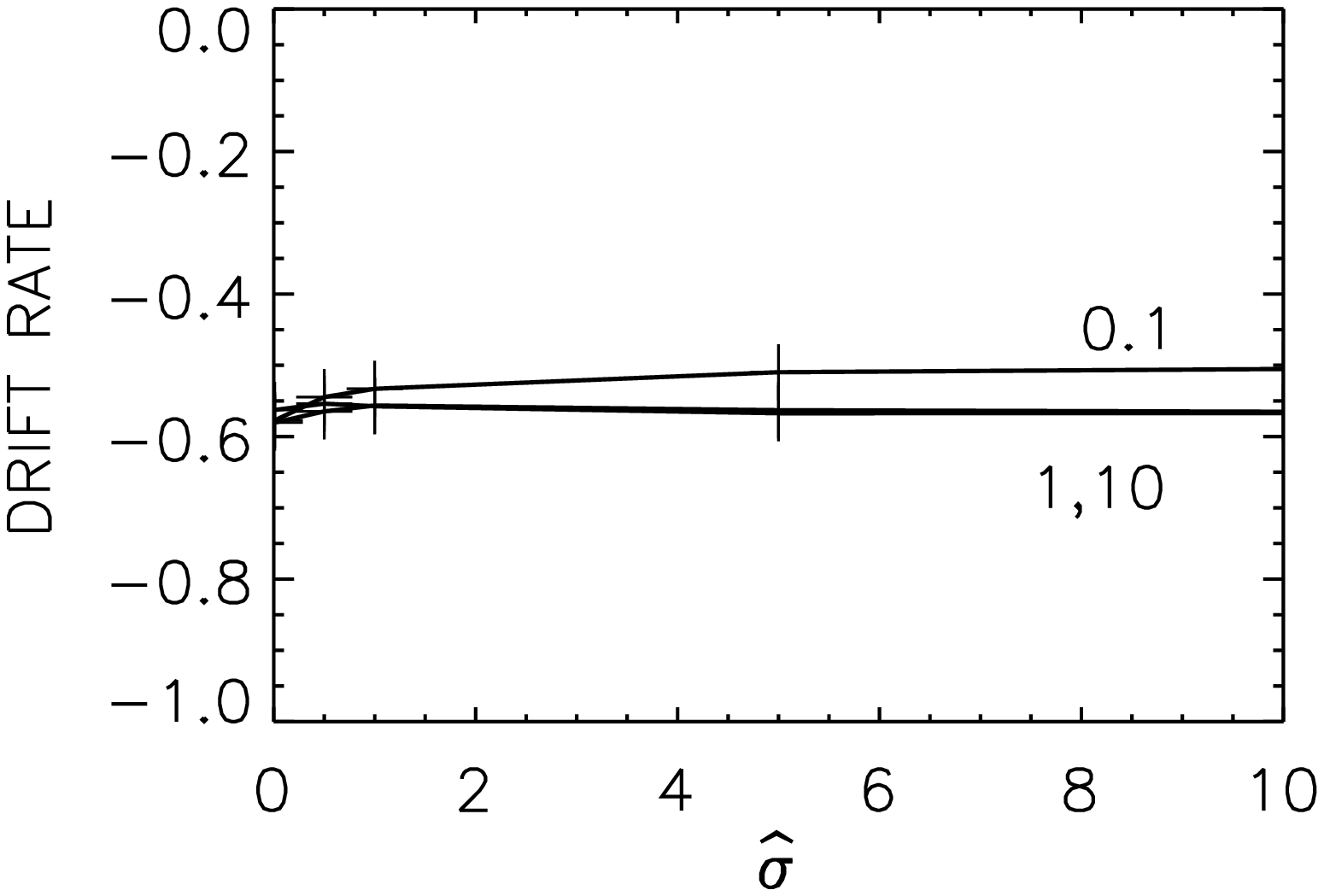}
}
 \caption{Normalized wave number $kR_0$(left) and drift frequency $\omega_{\rm dr}/m$ (right) of the marginal stability solutions shown in figure  \ref{fig3}. The curves are marked with the magnetic Prandtl numbers.} 
\label{fig4}
\end{figure}

The drift rates $\omega_{\rm dr}$ of the instability patterns are the real parts of the frequency $\omega$, normalized with the rotation rate of the inner cylinder.
Because of 
\begin{equation}
\frac{ \dot\phi}{\Om_{\rm in}}\,=\,-\,\frac{\omega_{\rm dr}}{m}\,,
\label{phi}
\end{equation}
the azimuthal migration $\dot\phi$ has the opposite sign as $\omega_{\rm dr}$. For AMRI with negative shear we always found that the pattern migrates for all $\Pm$ in positive $\phi$-direction. The right panel of figure  \ref{fig4} shows similar results. For all $\Pm$, negative $\omega_{\rm dr}$ occur and the perturbation pattern indeed co-rotates with the outer cylinder. For $\omega_{\rm dr}/ \Omin=-\mu $ the pattern would rotate just as the outer cylinder, hence the pattern migrates slightly faster than the outer cylinder rotates -- independent of the conductivity of the cylinders. The skin frequency $\omega/\Omin$ in the expression (\ref{az1}) does not depend on either $\hat\sigma$ or $\Pm$. 

This is not true for the wave numbers. They strongly depend  on the magnetic Prandtl number but again  they hardly depend on the conductivity ratio $\hat\sigma$. The wave numbers for the solutions with minimum magnetic field are extremal  for $\hat\sigma\simeq 0.5$. The waves are shorter for insulating or for perfectly conducting cylinders (figure  \ref{fig4}, left). The vertical extent $\delta z$ of the cells of the instability pattern normalized by the gap width $D$ is 
\beg
{\delta z \over D}\, = \,{\pi \over k R_0}
\sqrt{{\rin \over 1-\rin}}\,,
\label{delz}
\ende
hence $\rin=0.5$ leads to ${\delta z }/{ D} = {\pi}/{ k R_0}$, so that for $k R_0\simeq \pi$ the cells are almost square-shaped in the ($R/z$) plane. For $k R_0\gg\pi$ the cells are very flat. From the results given in figure  \ref{fig4} (left) the cells for small $\Pm$ are aligned with the rotation axis; they are most prolate for $\hats=0.5$. For $\Pm=10$ the cells are nearly square-shaped and the influence of $\hats$ almost vanishes.
\section{Potential flow, Rayleigh limit}\label{Potential}
A prominent example of AMRI is the potential flow with $\mu=\rin^2$, also known as the flow at the Rayleigh limit. The rotation profile is curl-free ($\ncurl~\vec{U}=0$), hence the name. This flow together with current-free azimuthal fields belongs to the Chandrasekhar-type MHD flows with $\vec{U}\propto \vec{B}$, for which it is known that their neutral-stability lines for $m=1$ converge for $\Pm\to 0$ in the ($\Ha/\Rey$) plane \citep{RS15}. This behavior is quite opposite to the scaling of the quasi-uniform flow with $U_\phi\simeq$~const discussed above. The viscosity of the fluid, that is, the magnetic Prandtl number, should influence the results more strongly than for the quasi-uniform flow. We shall also see that the different scalings for $\Pm\to 0$ do not lead to different consequences with respect to the influence of the finite conductivity of the cylinders.
\begin{figure}[ht]
 \centering
 \vbox{ 
 \includegraphics[width=0.49\textwidth]{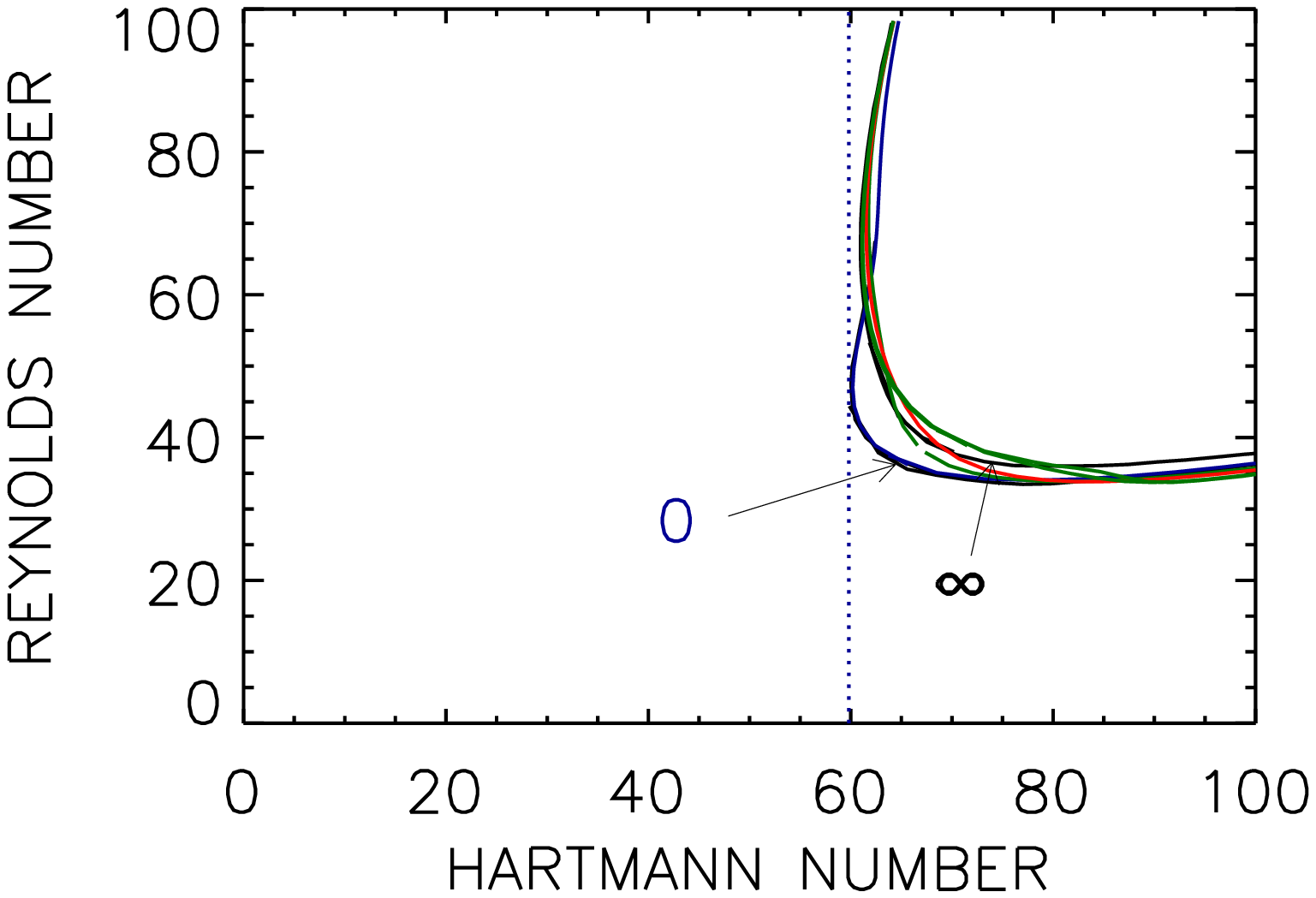}
 \includegraphics[width=0.49\textwidth]{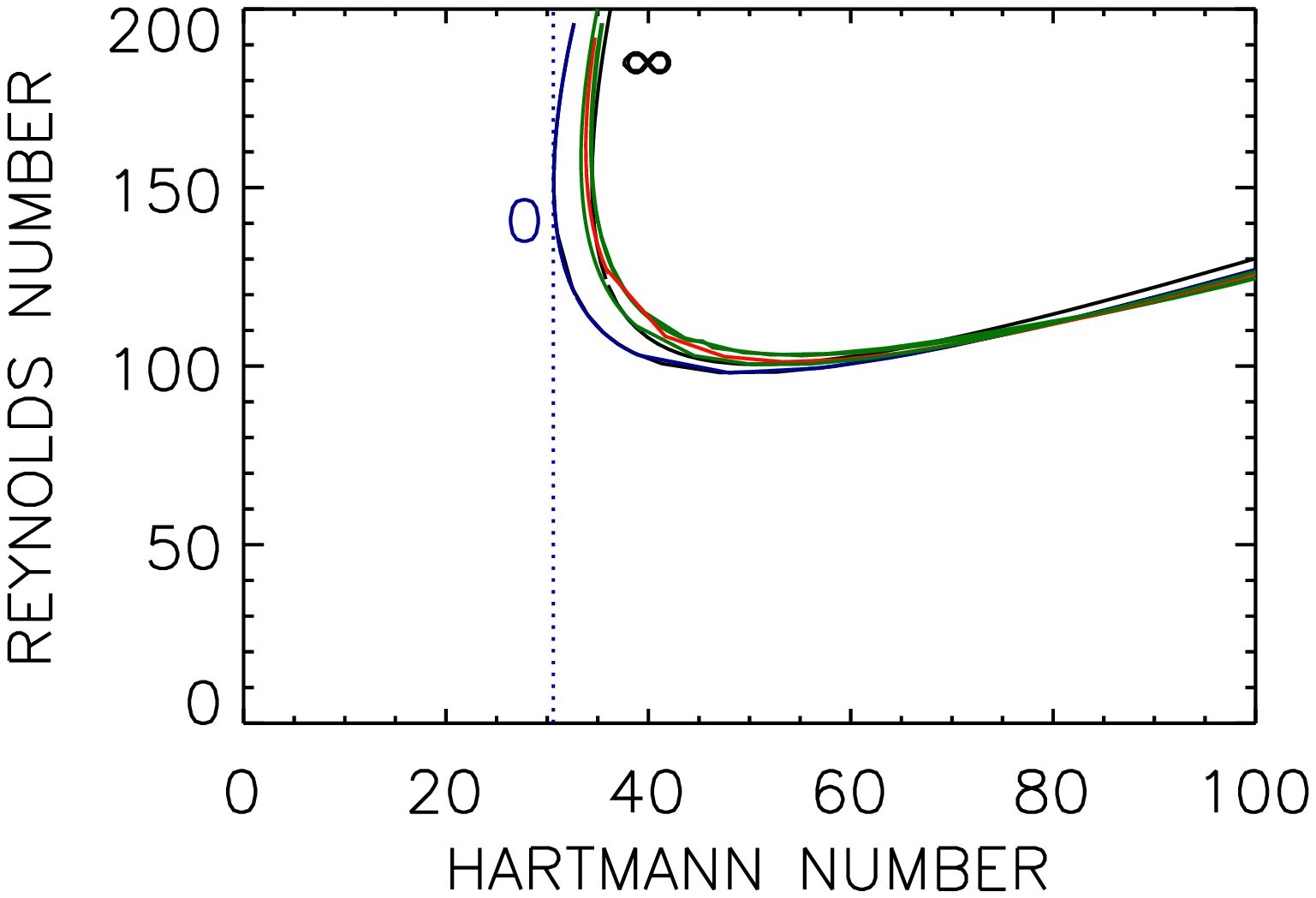}
}
 \caption{Stability maps for the potential flow and the current-free magnetic field between the cylinders. $\Pm=10$ (left), $\Pm=1$ (right). The curves are marked with their values of $\hat\sigma$: $\hats=\infty$ (black), $\hat\sigma=1$ (red), $\hat\sigma=0$ (blue), others (green). The vertical dotted line indicates the global minimum Hartmann number $\Ha_{\rm MIN}$. $m=1$, $\mu_B=\rin=0.5$, $\mu=0.25$. (colour online)} 
\label{fig10}
\end{figure}
\begin{figure}[ht]
 \centering 
 \includegraphics[width=0.49\textwidth]{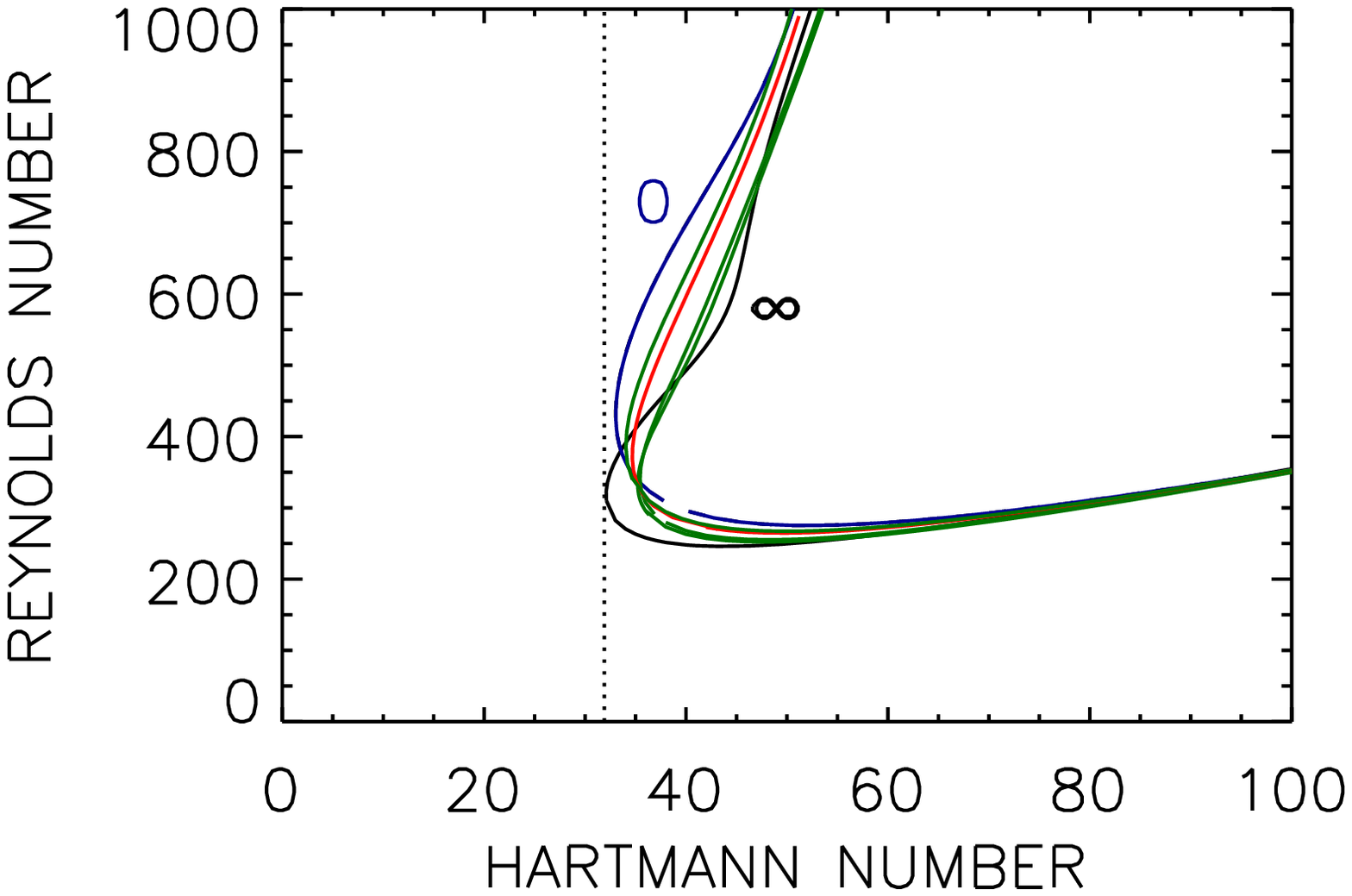} 
 \includegraphics[width=0.49\textwidth]{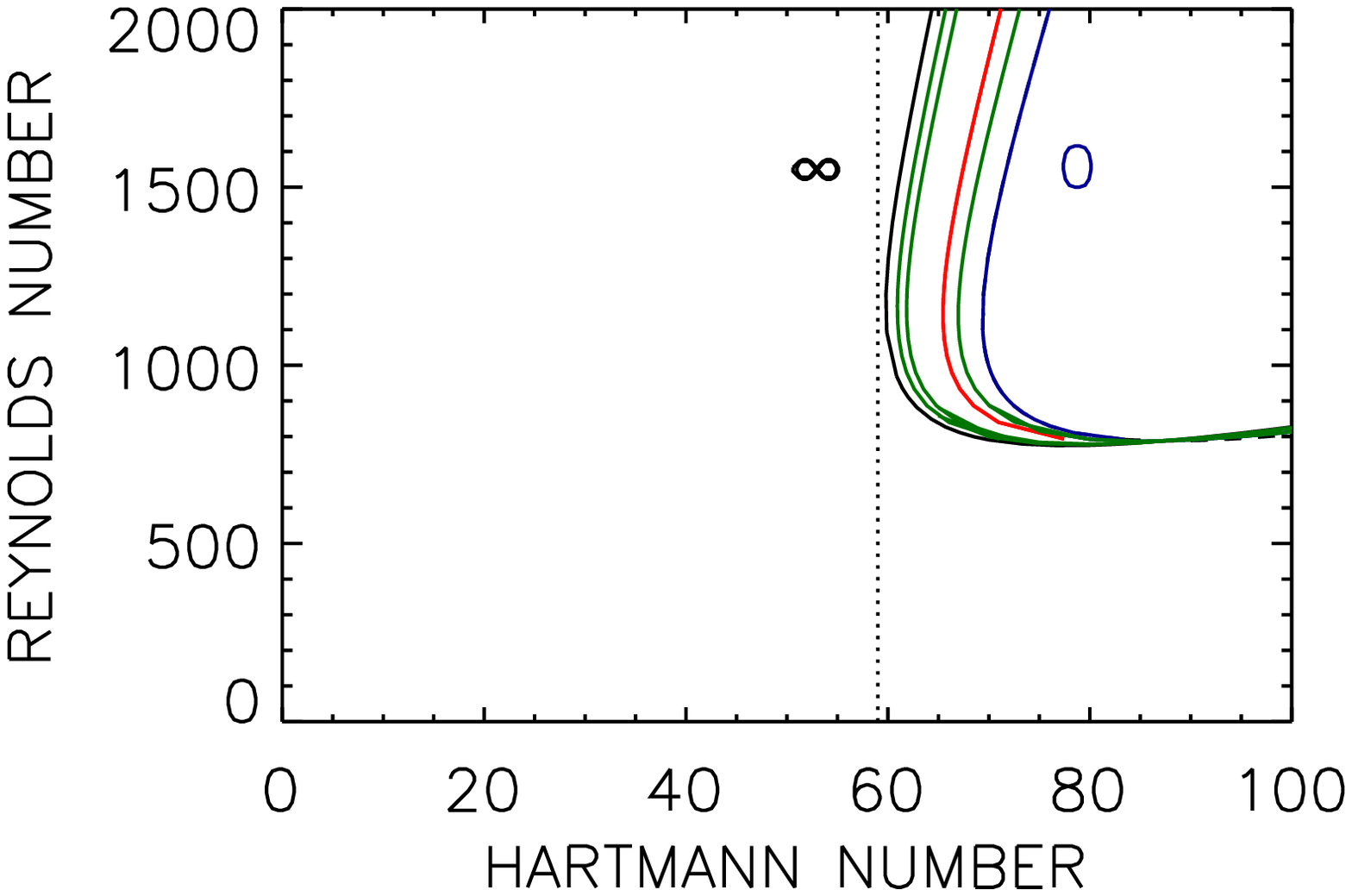}
 \caption{Similar to figure  \ref{fig10}, but for the small magnetic Prandtl numbers $\Pm=0.1$ (left) and $\Pm=10^{-5}$ (right). (colour online)}
\label{amri}
\end{figure}

Figures \ref{fig10} and \ref{amri} present the stability maps for large magnetic Prandtl numbers $\Pm\geq 1$, as well as the small magnetic Prandtl numbers $\Pm<1$ in the ($\Ha/\Rey$) plane, for various values of $\hat\sigma$. The map for $\Pm=10^{-5}$ is also valid for $\Pm\to 0$ \citep{HT10}. One again finds a characteristic influence of the magnetic Prandtl number, i.e. the slope of the function $\Ha_{\rm min}(\hats)$ at $\hats=0$ is negative for small $\Pm$ and positive for large $\Pm$. Note that the typical minimal magnetic Reynolds numbers in this plot are only $\rO(10^{-3}$), which shows that indeed the magnetic Prandtl number determines the form of $\Ha_{\rm min}(\hats)$, rather than the magnetic Reynolds number. Again for small $\Pm$ $\Ha_{\rm MIN}$ belongs to the perfectly conducting boundary conditions. For experiments with liquid metals with their low magnetic Prandtl numbers, therefore, the finite conductivity of the cylinders strongly influences only flows of high magnetic Reynolds numbers. 
\begin{figure}[h]
 \centering
\hbox{
 \includegraphics[width=0.49\textwidth]{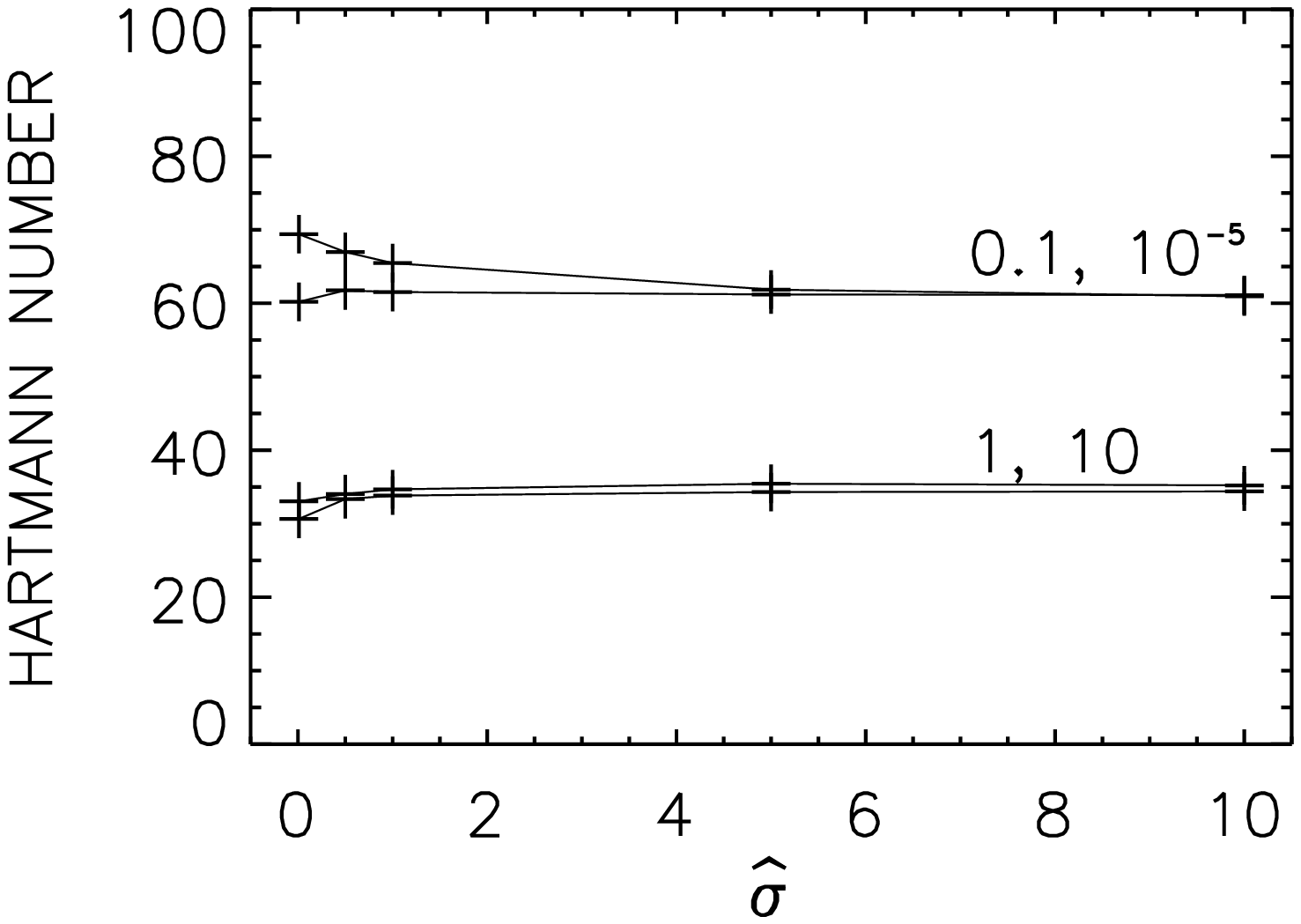} 
 \includegraphics[width=0.49\textwidth]{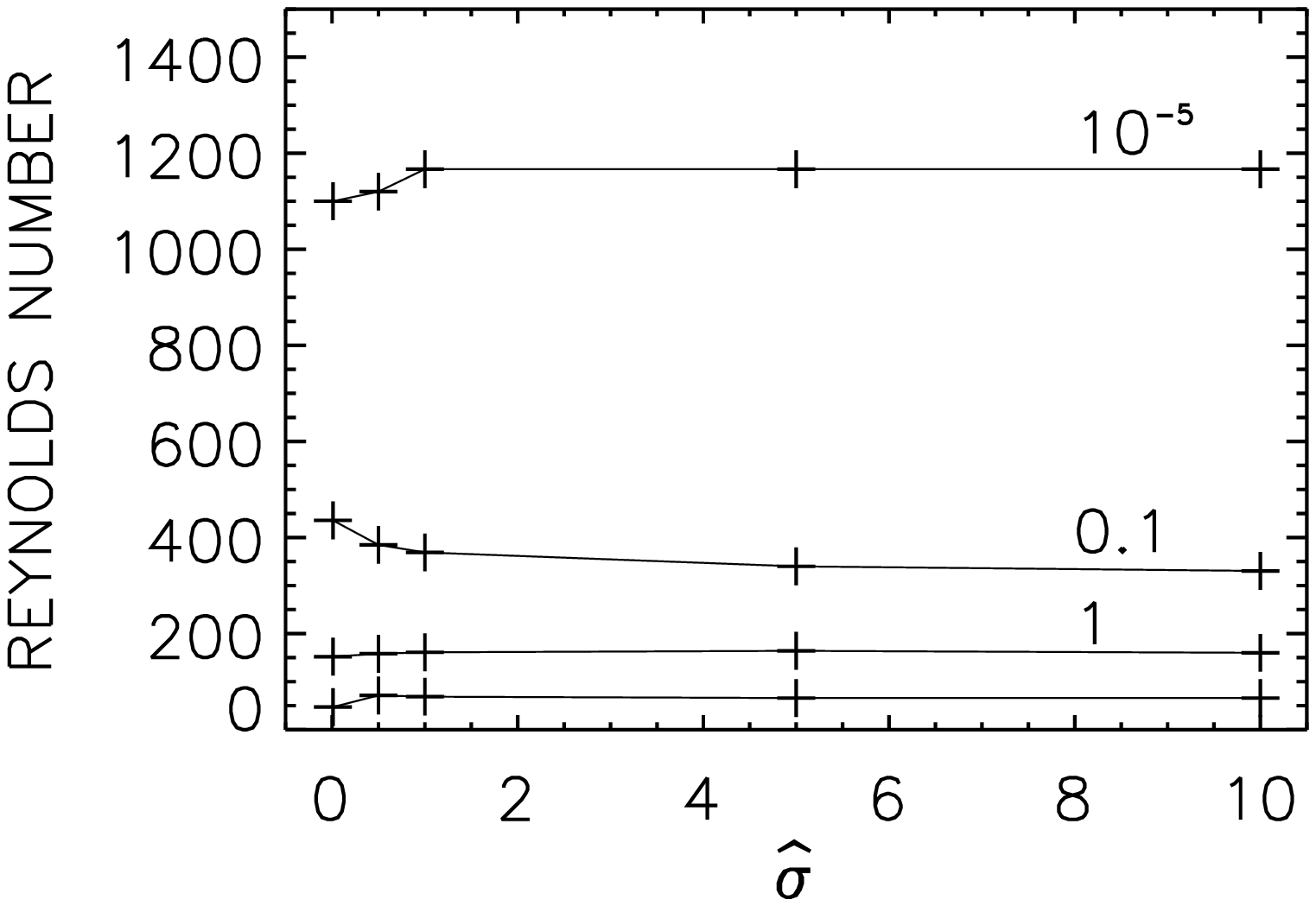} 
 }
 \caption{As in figure  \ref{fig3}, but for the potential flow with $\Pm=10$, $\Pm=1$, $\Pm=0.1$  and $\Pm=10^{-5}$ (marked). $m=1$, $\mu_B=\rin=0.5$, $\mu=0.25$.}
\label{fig12}
\end{figure}

With respect to the minimal Hartmann numbers for the instability onset, the results are rather similar to those of section \ref{Quasi}. For large $\Pm$ insulating boundaries yield $\Ha_{\rm MIN}$, while for small $\Pm$ perfectly conducting boundaries yield the minimum values. The transition happens for $\Pm\simeq 0.1$. The transition from the values $\hats=0$ (insulating walls) to $\hats=\infty$ (perfectly conducting walls) is always rather smooth (figure  \ref{fig12}). The magnetic Reynolds number is too small for more striking effects. 
\begin{figure}[h]
 \centering
\hbox{ 
 \includegraphics[width=0.49\textwidth]{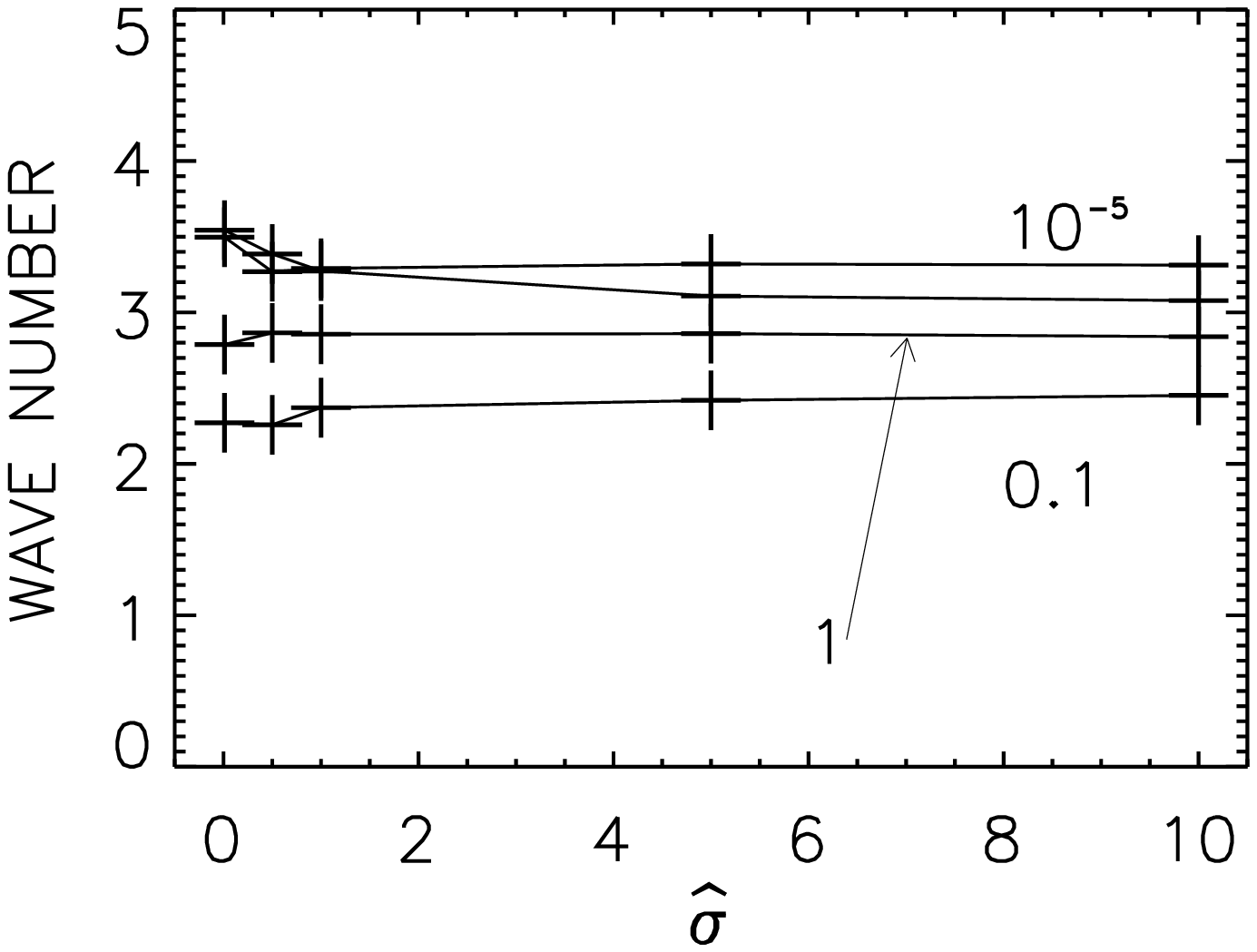}
 \includegraphics[width=0.49\textwidth]{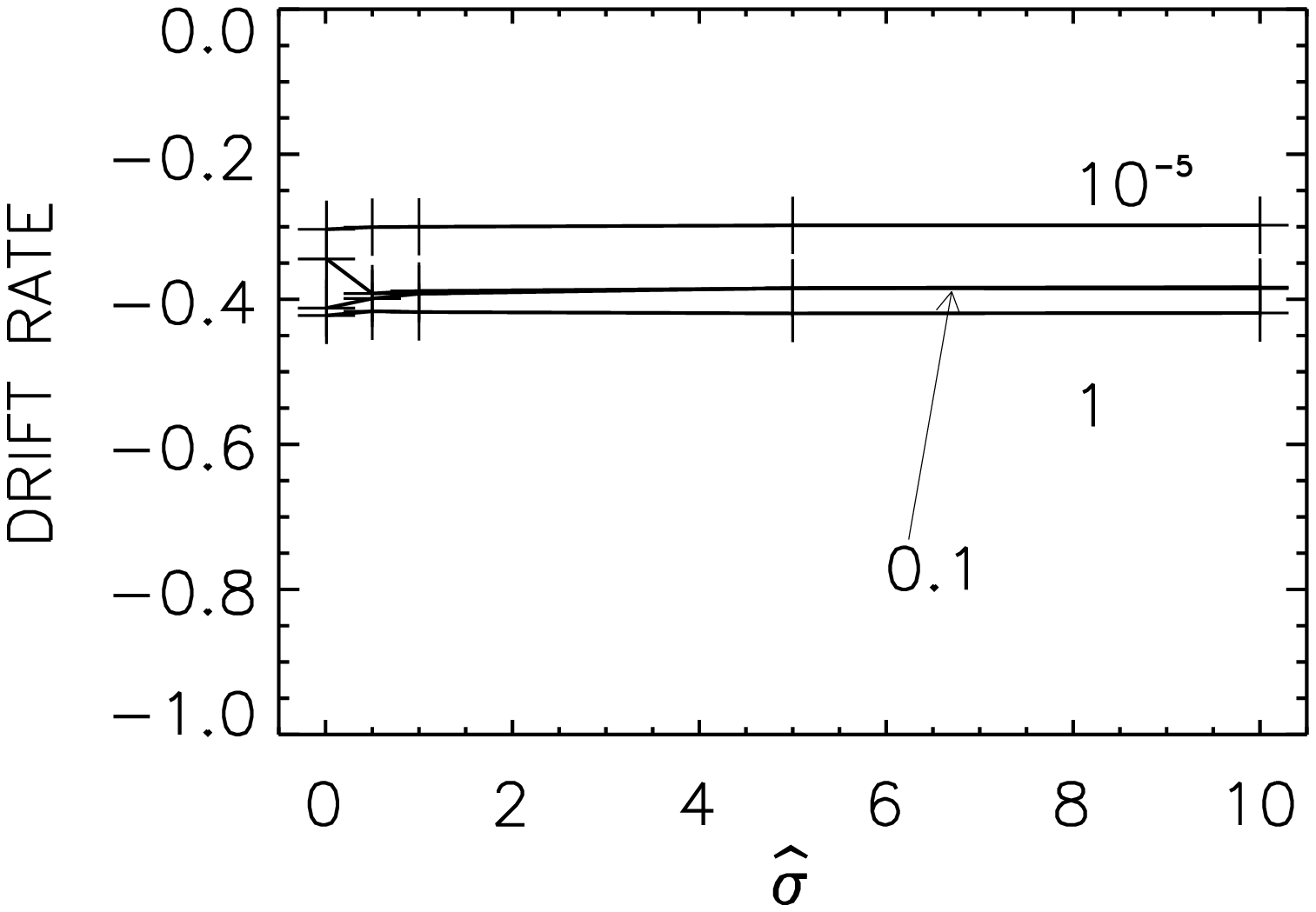}}
 \caption{Normalized wave number (left panel) and drift frequency $\omega_{\rm dr}$ (right panel) of the marginal stability solutions shown in figure  \ref{fig12} (same notations).}
\label{fig13}
\end{figure}

The normalized wave numbers and drift rates are given in figure  \ref{fig13}. For both cases the influence of the conductivity ratio $\hats$ is weak. The cells are almost-squarish in the ($R/z$) plane, and almost co-rotate with the outer cylinder. Exact co-rotation with the outer cylinder appears for $\dot\phi=\mu \Omin$, which is here only approximately  realized for small $\Pm$.
\section{Super-rotation}\label{Super}
Even rotation profiles with positive shear can be destabilized by current-free azimuthal fields \citep{SK15,RS16}. This is very surprising, as Taylor-Couette flows with positive shear seemed to be the prototype of hydrodynamic stability \citep{W33,SG59} -- but see \cite{D17} who found a linear hydrodynamic instability for very high Reynolds numbers. The hydromagnetic instability is a double-diffusive phenomenon which does not appear if the viscosity and the magnetic diffusivity are the same. It thus makes sense to discuss the stability problem for super-rotation only for small magnetic Prandtl number, here for $\Pm=10^{-5}$ as would be appropriate for experiments with liquid sodium.  Such experiments suggest themselves  to probe the spectacular destabilization of  super-rotating 
Taylor-Couette flows by simple current-free azimuthal magnetic fields. The first question
is  how the container must be constructed in order to reach the critical magnetic field strength in the laboratory. Only experiments with sufficiently strong fields  will show whether the marginally unstable nonaxisymmetric instability  patterns can be excited or not.  
\begin{figure}[h]
 \centering
 \includegraphics[width=0.65\textwidth]{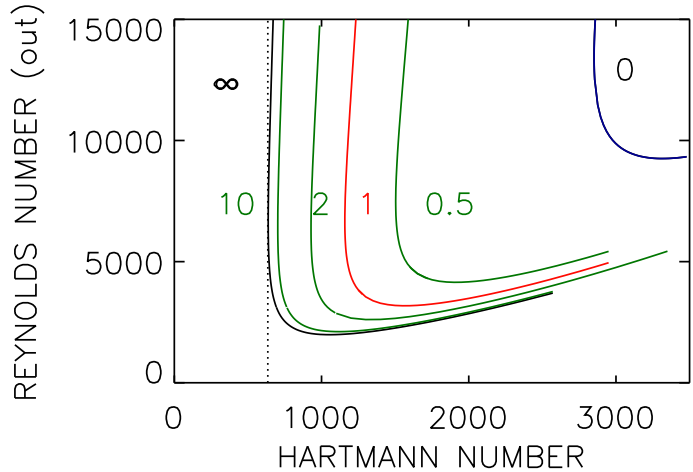}
 \caption{Stability maps for super-rotation subject to the current-free azimuthal magnetic field. The curves are marked with their values of $\hat\sigma$. The Reynolds number is formed with the outer rotation rate $\Om_{\rm out}$. The vertical dotted line indicates the global minimum Hartmann number $\Ha_{\rm MIN}$. $m= 1$, $\Pm=10^{-5}$, $\mu_B=\rin=0.9$, $\mu=5$. (colour online)}
\label{fig20}
\end{figure}

We start with the rotation profile $\mu=5$ in a narrow gap, where the outer cylinder rotates five times faster than the inner cylinder. In the following, for all rotation profiles with positive shear we shall use the {\em outer} Reynolds number
\begin{eqnarray}
 {\Rey}_{\rm out} \,=\,\frac{\Om_{\rm out} R_0^2}{\nu}=\mu \Rey\,,
\label{Reout}
\end{eqnarray} 
instead of the definition in (\ref{pm}). It is convenient also to define the drift rate (\ref{phi}) with respect to the outer rotation rate, i.e. $\omdr\to \omdr/\mu$. By this definition co-rotation of the instability pattern with the outer cylinder is described by $\omdr=-1$.
Figure \ref{fig20} presents the lines of neutral instability for $\hats=0\dots \infty$. The form of the resulting lines corresponds to the lines for AMRI with sub-rotation. For a given supercritical Reynolds number there is a minimum magnetic field for the onset of instability and a maximum magnetic field suppressing the instability again.

\begin{figure}[h]
 \centering
 \hbox{ 
 \includegraphics[width=0.49\textwidth]{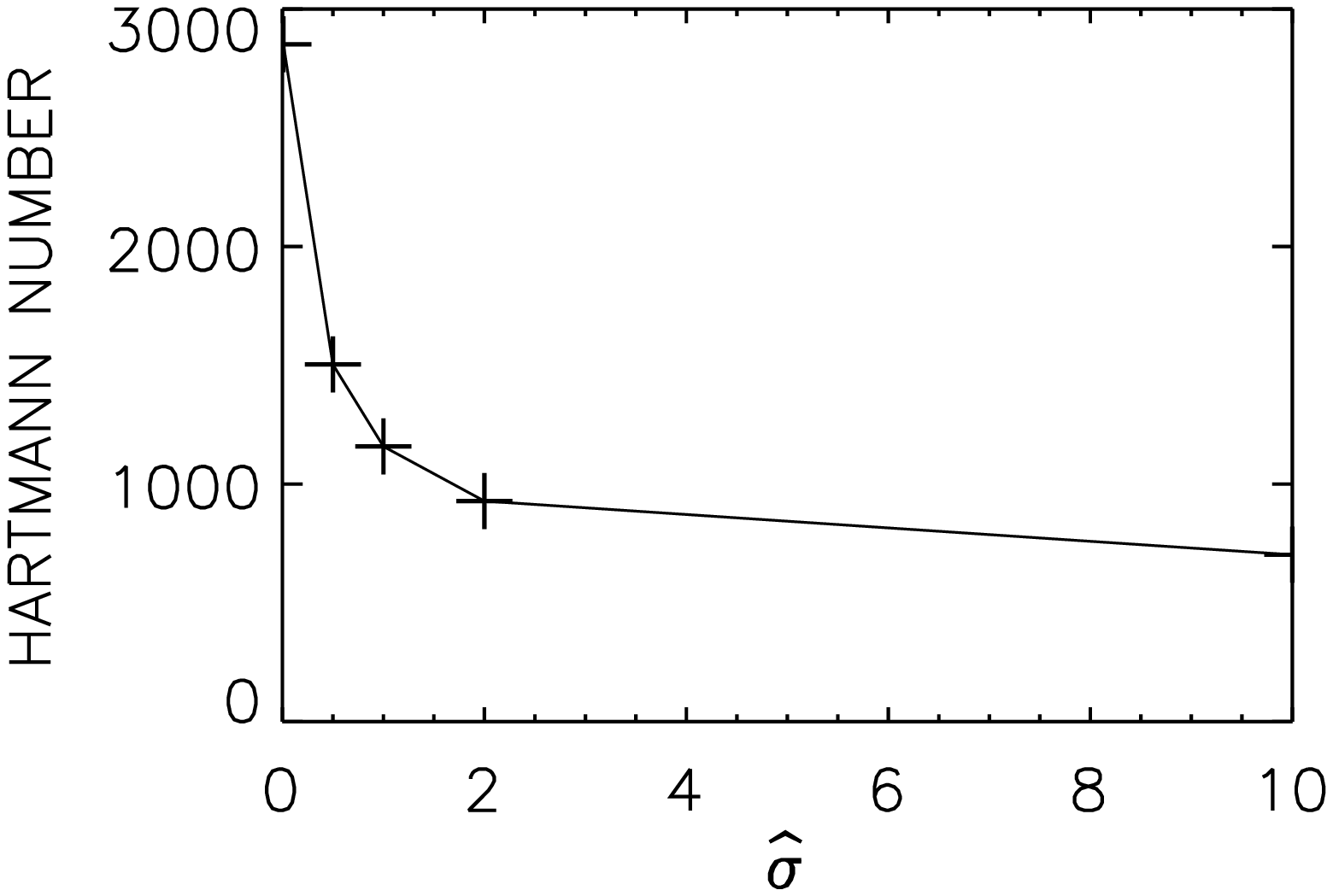} \hfill
 \includegraphics[width=0.49\textwidth]{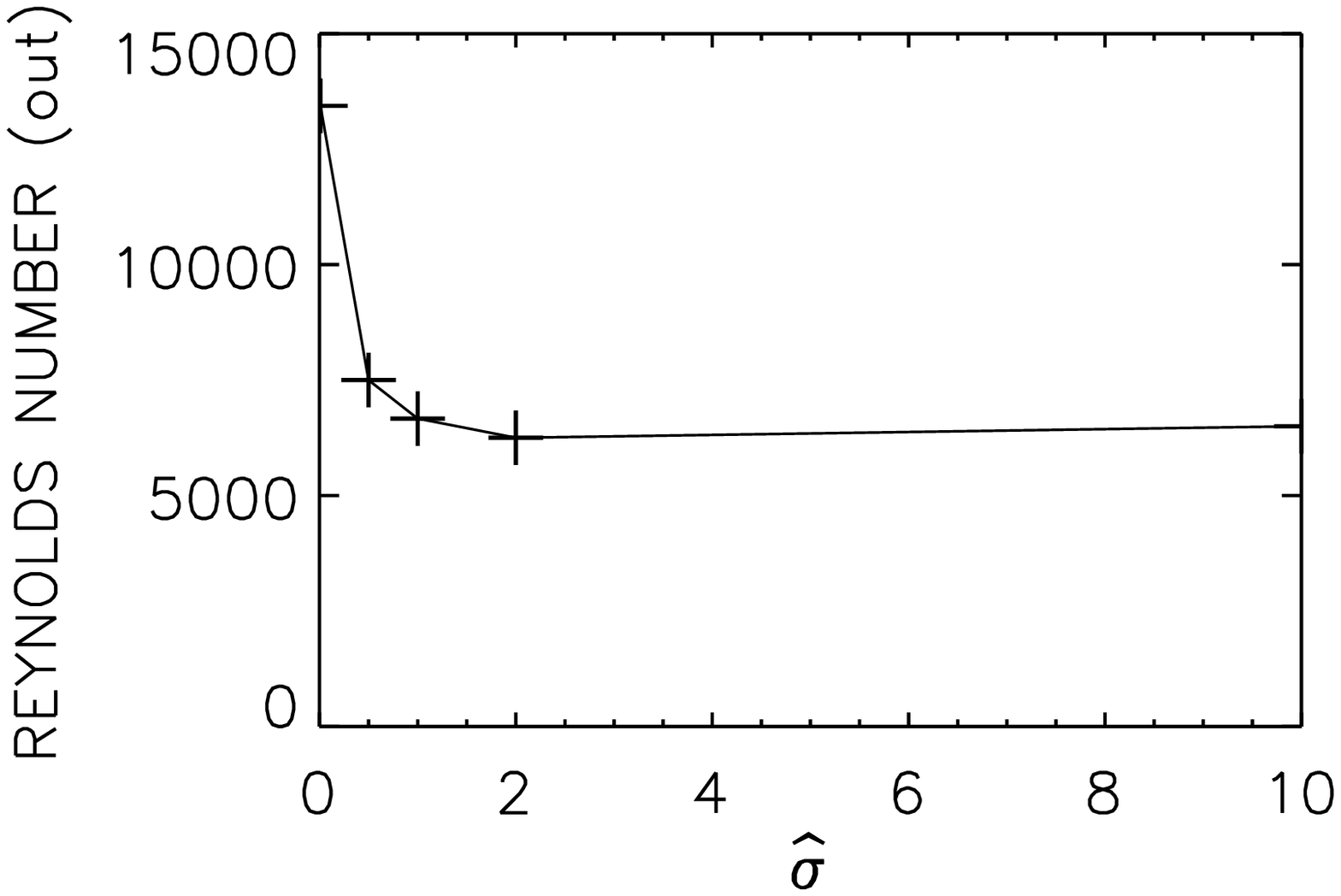} 
 }
 \caption{Same as figure  \ref{fig3} but for the super-rotation with $\mu=5$ in a narrow gap. $m= 1$, $\Pm=10^{-5}$, $\mu_B=\rin=0.9$.}
\label{fig21}
\end{figure}

One of the dramatic issues of the AMRI for super-rotation is the large separation of the lines for marginally stable solutions obtained for insulating and perfectly conducting boundary conditions. Figure \ref{fig21} shows a factor of more than three between the values of $\Hamin$ for $\hats=0$ and $\hats=\infty$. {\em All} curves for finite values of $\hats$ are located between the curves for insulating and perfectly conducting cylinders in a monotonic way; $\Hamin$ for $\hats\simeq 1$ approximately forms the median between the extreme values. For $\hats>1$ the function $\Hamin(\hats)$ becomes more and more a horizontal line. One thus finds the lines for $\hats\simeq 5$  already close to the absolute minimum value $\HaMIN\simeq 750$ valid for perfectly conducting cylinders. The influence of the finite conductivities of the cylinders, therefore, is {\em not} too essential for this flow. On the other hand, as $\Hamin(\hats)$ for $\hats<1$ scales as $1/\sqrt{\hats}$ the inverse conductivity ratio 1/5 provides a minimum Hartmann number close to that for insulating boundary conditions.

\section{Stationary inner cylinder}\label{stationary}
Taylor-Couette flows with stationary inner cylinder are clearly the most basic  models for super-rotating fluids. There is a long history to probe their hydrodynamic stability or instability. Interestingly enough, it has been shown by numerical simulations for very high Reynolds numbers that a hydrodynamic instability does exist \citep{D17}. In a previous paper we have shown that under the influence of current-free azimuthal fields such flows become unstable even for much lower Reynolds numbers of order $10^3$. The results, however, are only valid for narrow gaps between the cylinders, with the consequence that the solutions for perfectly conducting and insulating cylinders differ significantly. Only for perfectly conducting cylinders is $\Hamin$ small enough for realistic experiments  \citep{RS19}.
\subsection{Narrow gap}
\begin{figure}[h]
 \centering
 \includegraphics[width=0.65\textwidth]{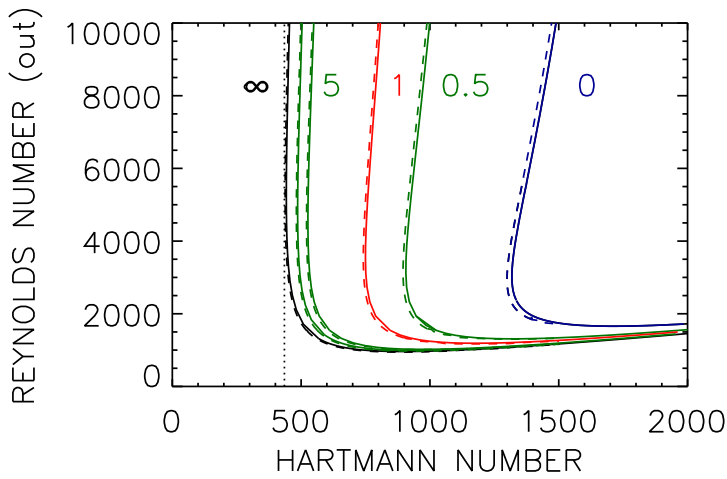}
 \caption{Stability maps for Taylor-Couette flows with almost stationary inner cylinders subject to the current-free azimuthal magnetic field for $\mu=128$ (solid lines), and $\mu=512$ (dashed lines). The Reynolds numbers are formed with the outer rotation rate $\Om_{\rm out}$. The curves are marked with their values of $\hat\sigma$.  The vertical dotted line indicates the global minimum Hartmann number $\Ha_{\rm MIN}$. $m= 1$, $\Pm=10^{-5}$, $\mu_B=\rin=0.9$. (colour online)}
\label{fig30}
\end{figure}

One question is whether the $\Hamin$ value for realistic $\hats$ is close enough to this solution. As in the foregoing paper we are forced to model stationary inner cylinders described by $\mu\to\infty$ which the code is able to approximate with very high $\mu$-values. Figure \ref{fig30} presents the instability maps for $\mu=128$ and $\mu=512$ for $\Pm=10^{-5}$ and for various $\hats$ in the $(\Ha/\Rey)$ plane. Obviously, the model with $\mu=128$ already gives an excellent approximation for the rotation profile with stationary inner cylinder. As expected for small $\Rm$ the absolute minimum $\HaMIN$ of the critical Hartmann number corresponds to the {\em perfectly conducting} boundary condition. One also finds that the line for $\hats\simeq 1$ lies in the middle of the instability domain defined by the $\Hamin$ of the two extremes perfectly conducting and insulating cylinders. 

\begin{figure}[h]
 \centering
 \hbox{ 
 \includegraphics[width=0.49\textwidth]{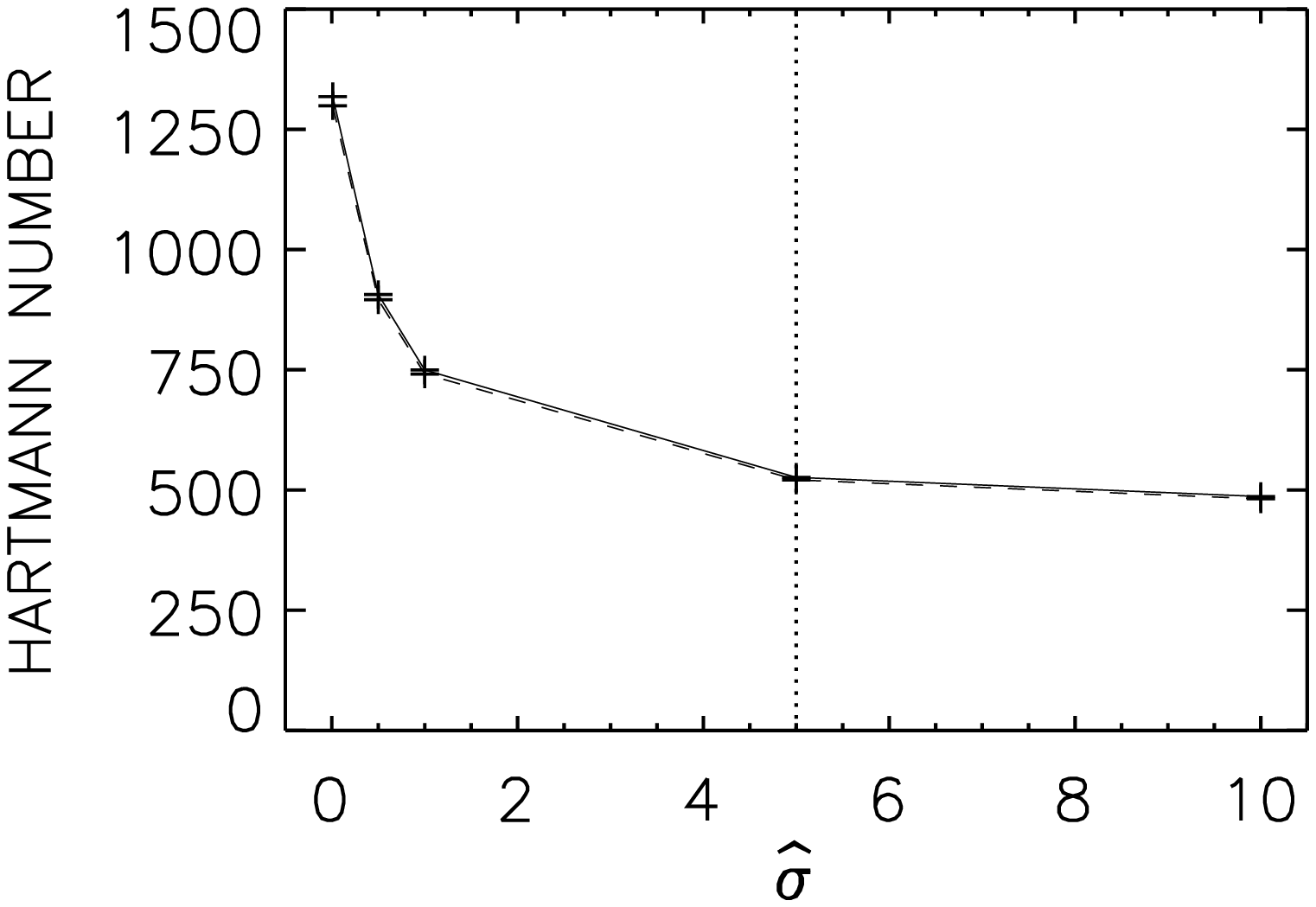} \hfill
 \includegraphics[width=0.49\textwidth]{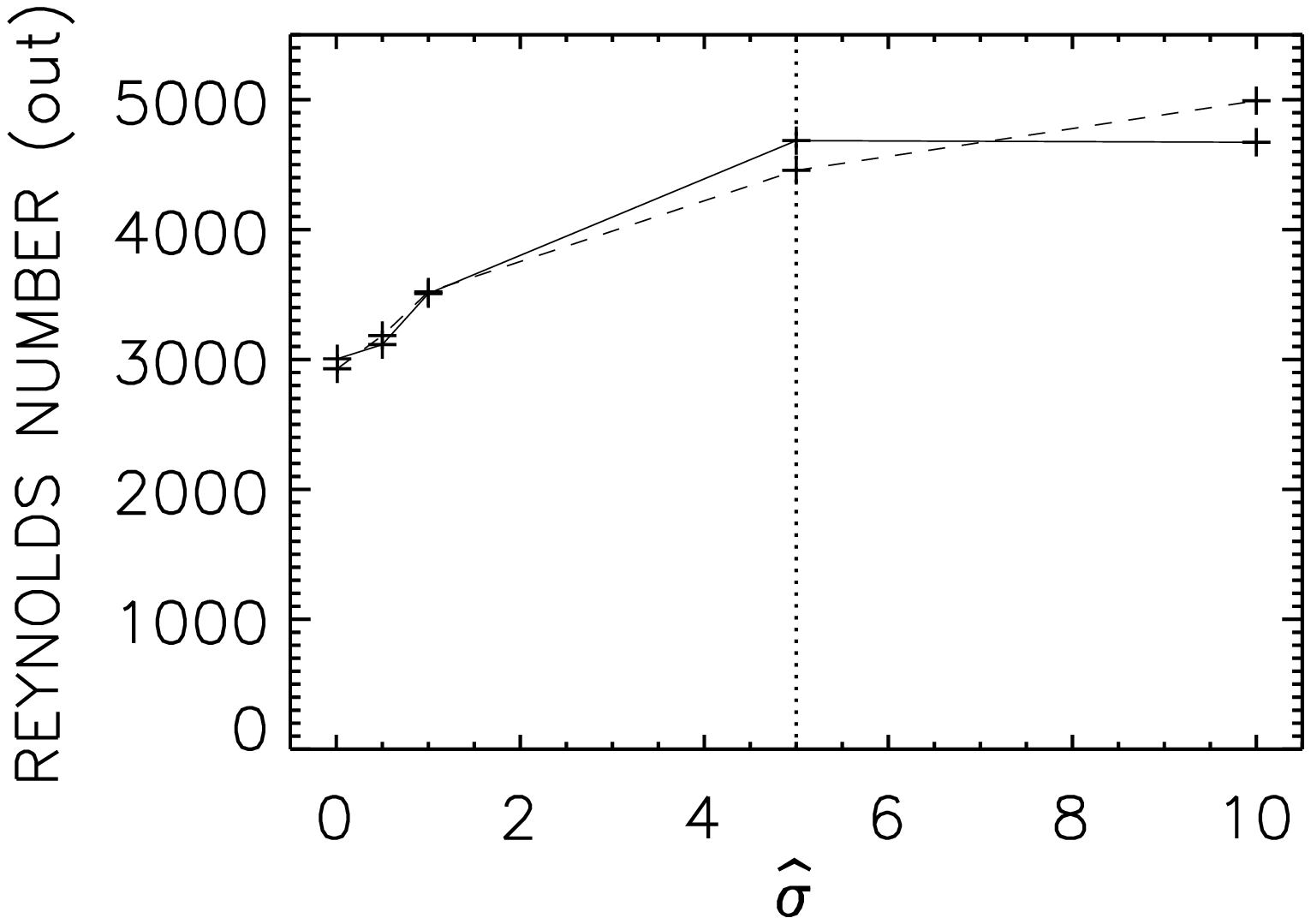} 
 }
 \caption{Similar to figure  \ref{fig3}, but for rotation with stationary inner cylinder in a narrow gap. $m= 1$, $\Pm=10^{-5}$, $\mu_B=\rin=0.9$, $\mu=128$ (solid line), $\mu=512$ (dashed line) The vertical dotted lines mark the position of $\hats=5$.
 }
\label{fig31}
\end{figure}

The onset values $\Hamin$ as a function of $\hats$ together with the associated Reynolds numbers are given by the plots in figure  \ref{fig31}. One finds a characteristic structure of this function. It is rather flat as a function of $\hats$ for $\hats>1$ and is also flat for $\hats<1$ as a function of $1/\hats$. The consequence is that the solutions for (say) $\hats>5$ are located close to the line for $\hats=\infty$, and the solutions for (say) $\hats<1/5$ are located close to the line for $\hats=0$. To minimize the required magnetic field  it is thus necessary that the conductivity of the cylinder material exceeds the conductivity of the liquid metal by a factor of (say) five. Both the minimum Hartmann number and the associated Reynolds number for $\hats>5$ only differ slightly from the values valid for perfect conductors. Only a small amplification is necessary for the electric current inside the inner cylinder to probe the stability of super-rotating liquid sodium under the influence of azimuthal magnetic fields if realistic boundary conditions are applied. 


Figure \ref{fig30} also demonstrates that $\HaMIN$ for very large $\mu$ is much smaller than for $\mu=5$. The instability for azimuthal current-free field with stationary inner cylinder for very small $\Pm$ is thus shown to be that form of  magnetorotational instability which is easiest to excite in the laboratory. 

\begin{figure}[h]
 \centering
 \vbox{ 
 \includegraphics[width=0.49\textwidth]{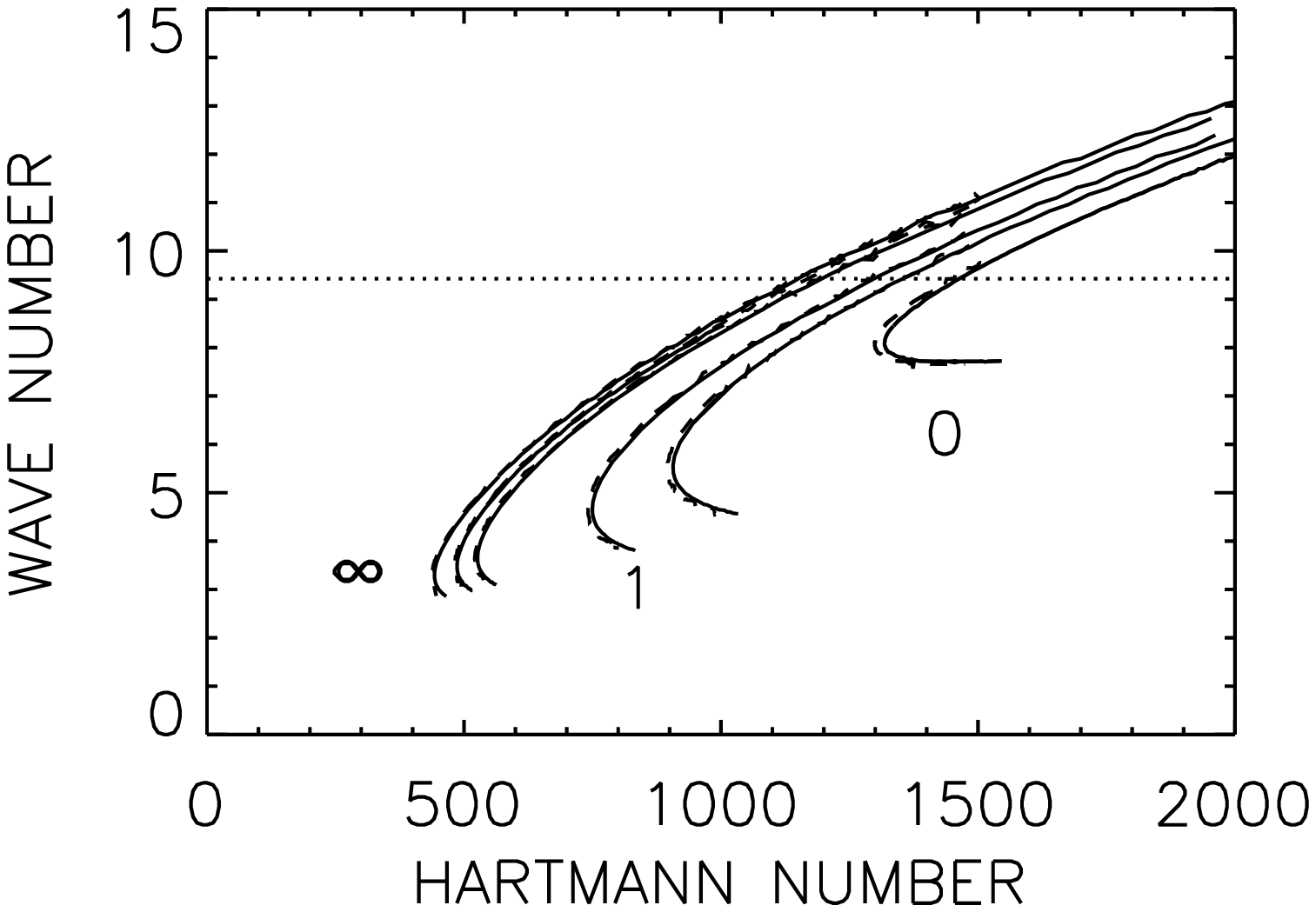} 
 \includegraphics[width=0.49\textwidth]{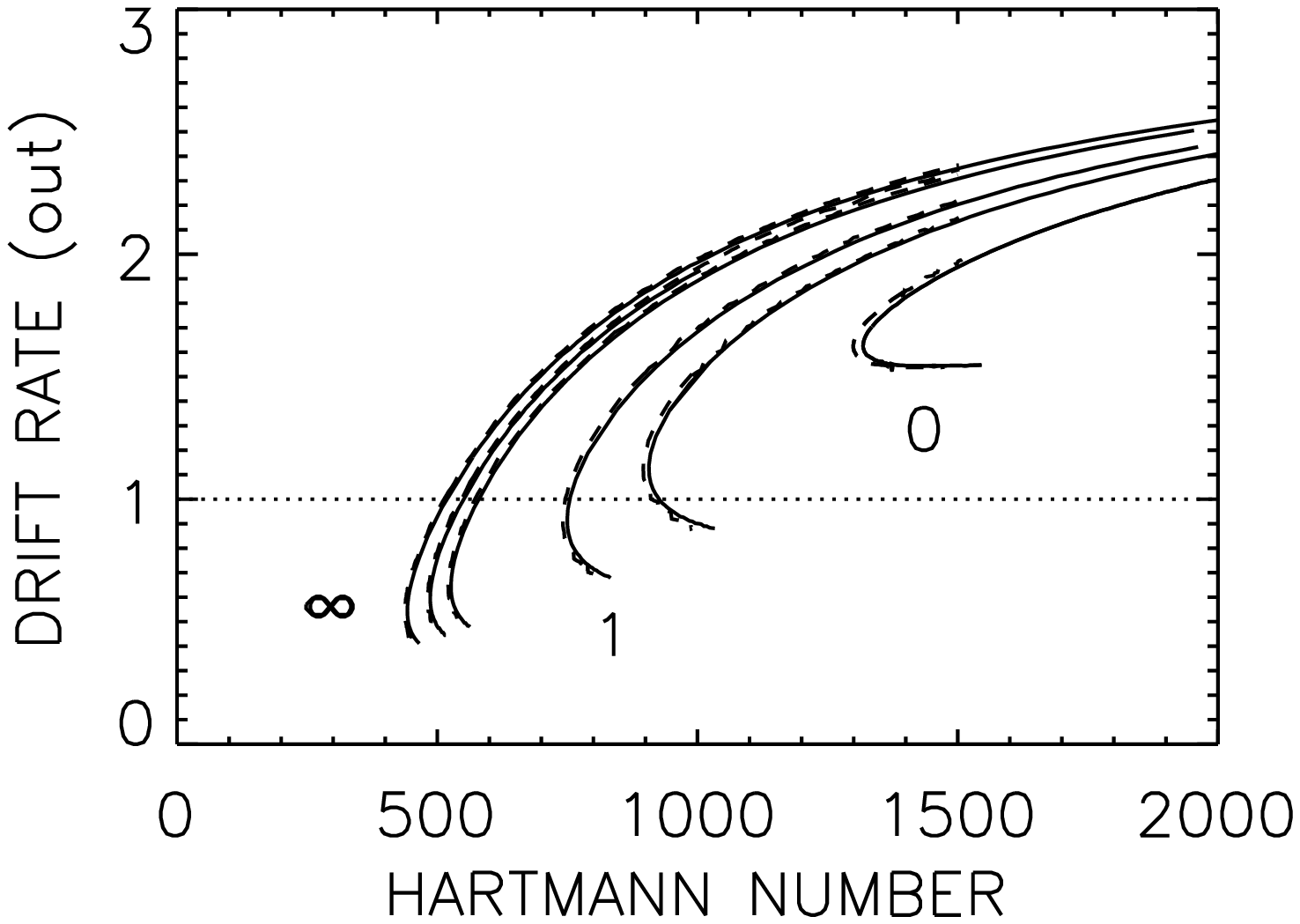} 
 }
 \caption{Similar to figure  \ref{fig30} (stationary inner cylinder, $\mu=128$) but for the normalized wave number $kR_0$ (left) and the drift frequency normalized with  the outer rotation rate $\Om_{\rm out}$ (right). The horizontal lines mark the limit of squarish cells and exact counter-rotation with the outer cylinder. }
\label{fig30b}
\end{figure}

The wave numbers in figure  \ref{fig30b} must be interpreted in light of Eq. (\ref{delz}) which leads for $\rin=0.9$ to ${\delta z }/{ D} = {3 \pi}/{ k R_0}$, so that for $k R_0\simeq 3 \pi$ (dotted line) the cells are almost square in the ($R/z$) plane. The cells described by the left panel of figure  \ref{fig30b} are all oblong with respect to the rotation axis. The strong influence of the boundary conditions on the shape of the instability cells (a factor of two!) is also puzzling.

The drift rates for stationary inner cylinder, normalized with the rotation rate of the outer cylinder, are given by the right panel of figure  \ref{fig30b} taken along the stability lines of figure  \ref{fig30}. Given only for the minima of the Hartmann number of each curve, the drift rate varies between 0.6 and 1.5 for the two extreme boundary conditions, approaching unity for $\hats\simeq 1$. Hence, for super-rotation and for $\hats\simeq1$, the instability pattern counter-rotates while for sub-rotation it co-rotates with the outer cylinder, in both cases with $|{\dot\phi}|=\Omout$. Only these two possibilities appear to exist for small $\Pm$. One finds that for small $\Pm$ and for $\hats\simeq 1$ the azimuthal drift equals the rotation of the outer cylinder but with opposite signs for sub-rotation and super-rotation. If the sign of the shear is changed (not the sense of rotation!) in experiments with $\Pm\ll1$ the direction of the drift simultaneously changes. { The different signs of the drift rates for sub-rotation and super-rotation indicate the wide difference of the nonaxisymmetric instabilities for the different rotation laws. Note that it can be rather misleading to interprete the azimuthal drift of the instability pattern as the basic rotation of the flow.} 

\subsection{Gap width variations}
\begin{figure}[h]
 \centering
 \vbox{ 
 \includegraphics[width=0.49\textwidth]{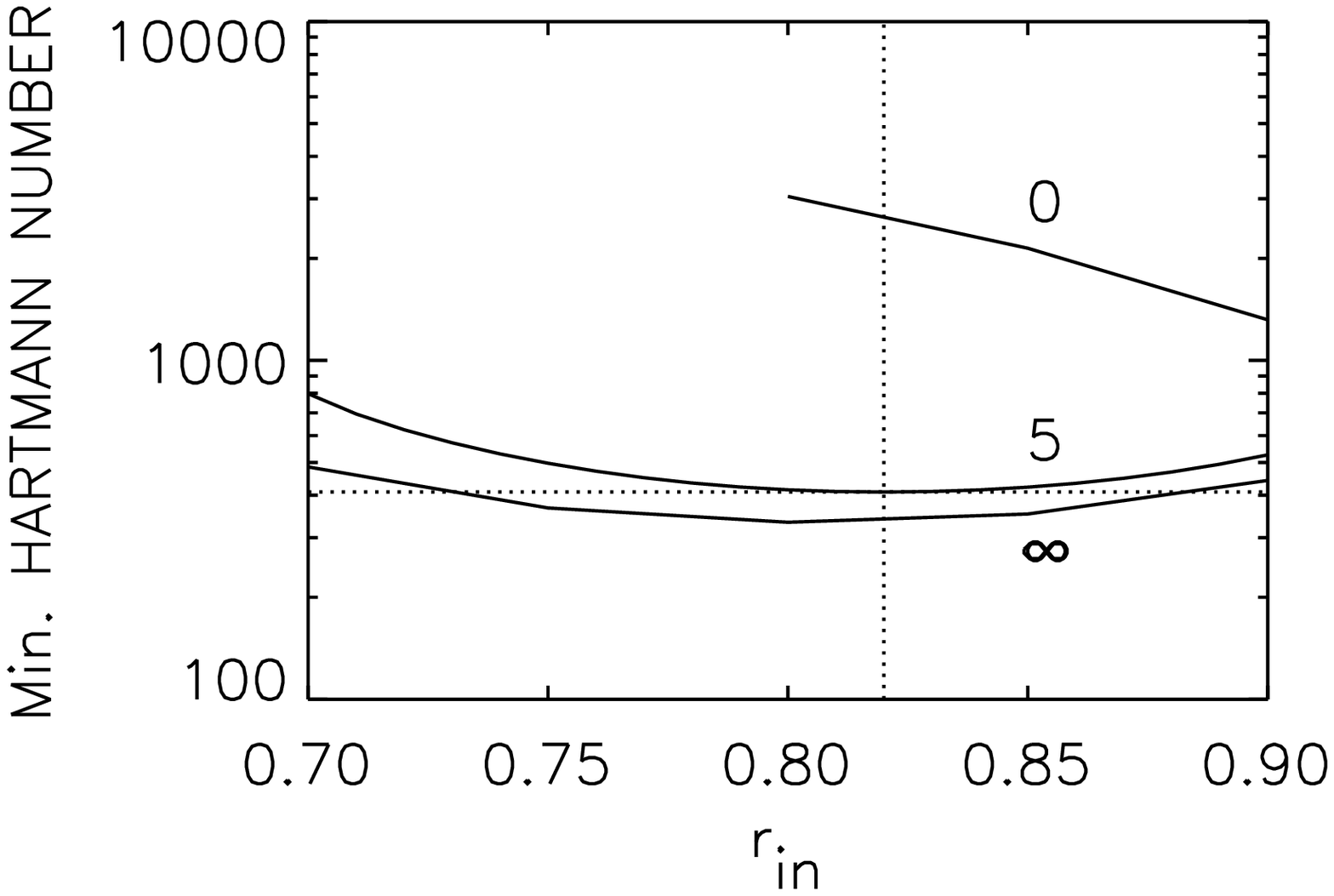} 
 \includegraphics[width=0.49\textwidth]{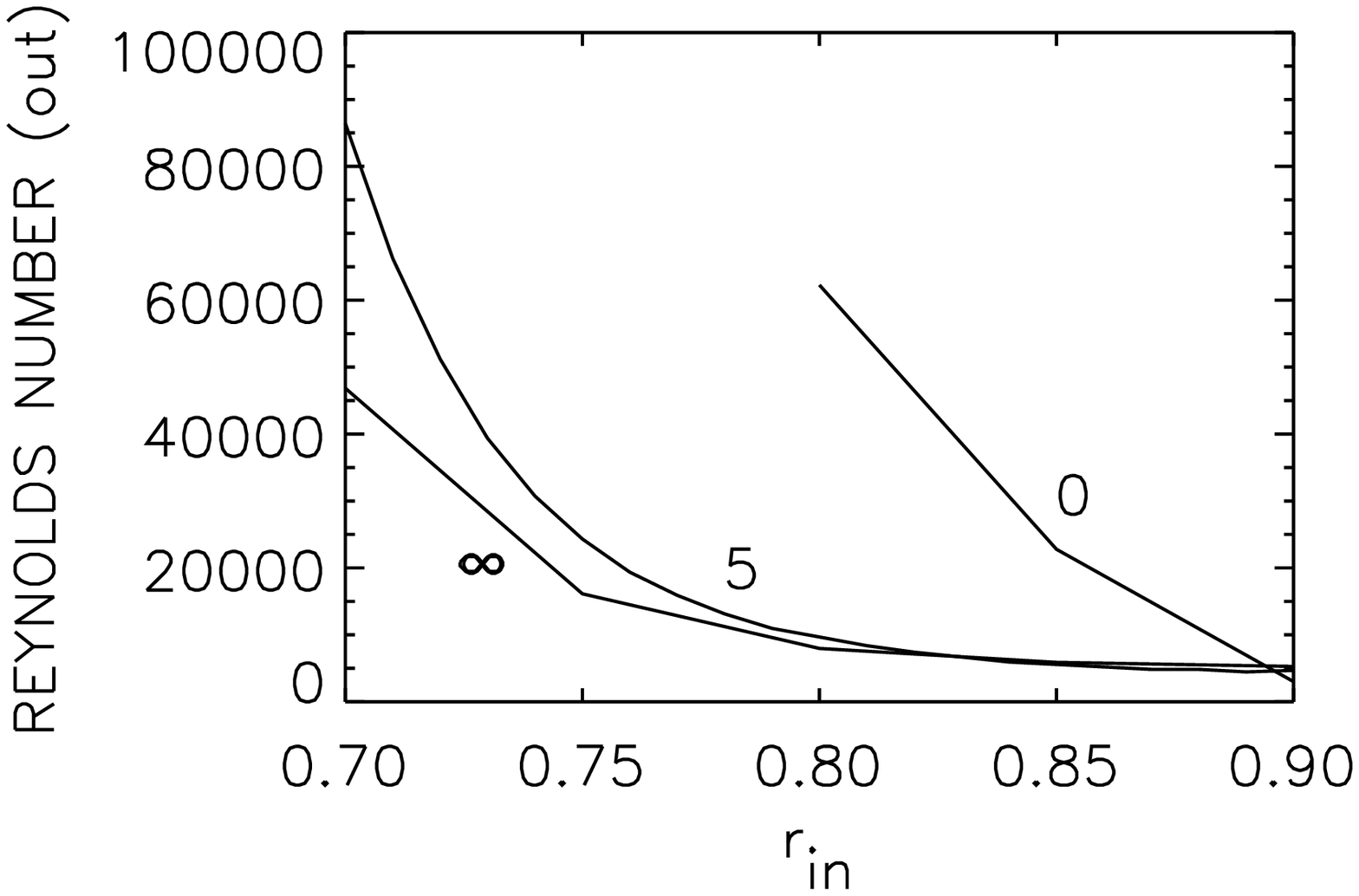} 
 }
 \caption{Minimum Hartmann numbers $\Hamin$ for the onset of the instability (left) and associated  Reynolds numbers (right) for various gap widths. $\hats=0$, $\hats=5$     and $\hats=\infty$ (marked). $m= 1$, $\Pm=10^{-5}$, $\mu_B=\rin$, $\mu=128$. The dotted lines in the left panel mark the position of the absolute minimum of $\Hamin$ for $\hats=5$.}
\label{gaps}
\end{figure}

\begin{figure}[h]
 \centering
\includegraphics[width=0.65\textwidth]{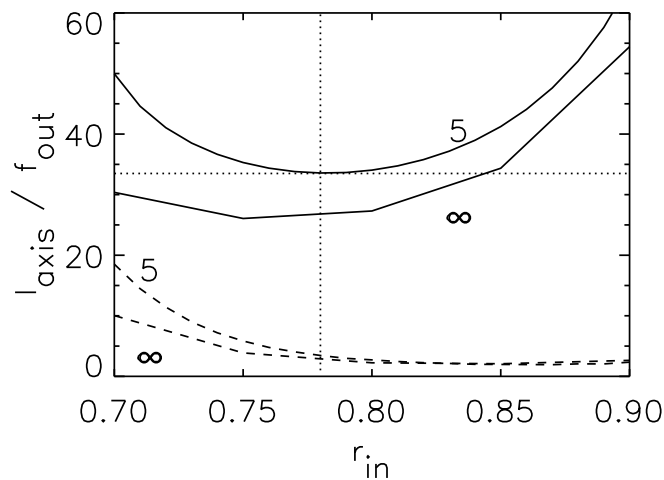} 
 \caption{Minimum electric current in kA (solid lines) and frequency in Hz of the outer cylinder (for $\Rout=5$~cm, dashed lines) for the onset of the instability for various gap widths and for liquid sodium as the fluid conductor. $\hats=5$  and $\hats=\infty$. $m= 1$, $\Pm=10^{-5}$, $\mu_B=\rin$, $\mu=128$  The dotted lines mark  the location of the absolute minimum of the critical electric current for $\hats=5$.}
\label{currentmin}
\end{figure}

It remains to vary the gap width for the flows with stationary inner cylinders. In order to transform the numerical values of the Hartmann numbers to axial electric currents (within the inner cylinder) the relation $I_{\rm axis}=5 R_{\rm in} B_{\rm in}$ may be written as
\begin{eqnarray}
 I_{\rm axis}\,=\, 5 \sqrt{ \frac{\rin}{1-\rin} } \ \Ha\ \sqrt{\mu_0\rho\nu\eta}\,,
\label{current} 
\end{eqnarray}
with $\sqrt{\mu_0\rho\nu\eta}\simeq 8.2$ in cgs units for liquid sodium. The axial currents are measured in Ampere, the radius in cm and the magnetic fields in Gauss. According to (\ref{current}) the Hartmann number $\Hamin=442$ for stationary inner cylinder with $\rin=0.9$ and for perfectly conducting boundaries transforms to an electric current of $I_{\rm axis}=55$~kA, which would require enormous experimental effort. The question is whether this value is reduced by increasing the gap width. Figure \ref{gaps} gives $\Hamin$ and the corresponding Reynolds numbers for $0.7 \leq \rin\leq 0.9$ for insulating cylinders, perfectly conducting cylinders and for $\hats=5$. The latter two lines show local minima for $\rin\simeq 0.82$. For this gap width the finite conductivity of the cylinders ($\hats=5$) enhances $\Hamin$ only from 340 to 409. However, the local minimum for the axial electric current which generates the azimuthal magnetic field appears for slightly wider gaps. From figure  \ref{currentmin} one finds the minimum of the axial electric current for $\hats=5$ at $\rin=0.78$ as $I_{\rm axis}=33.5$~kA. For perfectly conducting cylinders this value would reduce to $I_{\rm axis}=27$~kA so that the relative increase of the critical electric current by finitely conducting cylinders is about 25\%.

From the Reynolds numbers the frequency of the outer cylinder can be determined if the size of the container is known. The typical viscosity of liquid sodium is $7\cdot 10^{-3}$cm$^2$/s. For (say) $\Rout=5$~cm one obtains
\begin{eqnarray}
 f_{\rm out}\,= \,4.5\cdot 10^{-5} \frac{\Rey_{\rm out}}{\rin(1-\rin) } \ \ \ \ \ \ \ {\rm [Hz]}\,,
\label{frequency} 
\end{eqnarray}
resulting in about 3~Hz for the critical frequency of the outer cylinder for perfectly conducting cylinders and  $\rin=0.9$ (as an example),  and nearly the same value for $\hats=5$ (see figure  \ref{currentmin}).

\subsection{Eigenfunctions}
The eigenfunctions $\vec{u}(R)$ and $\vec{b}(R)$ of the linearized equations can be computed with the known eigenvalues for neutral instability, but only up to multiplication by an arbitrary real factor. This factor does not influence their zeros, and the sign of products (or ratios)  of two  perturbation components also remains unchanged.
\begin{figure}[h]
 \includegraphics[width=0.33\textwidth]{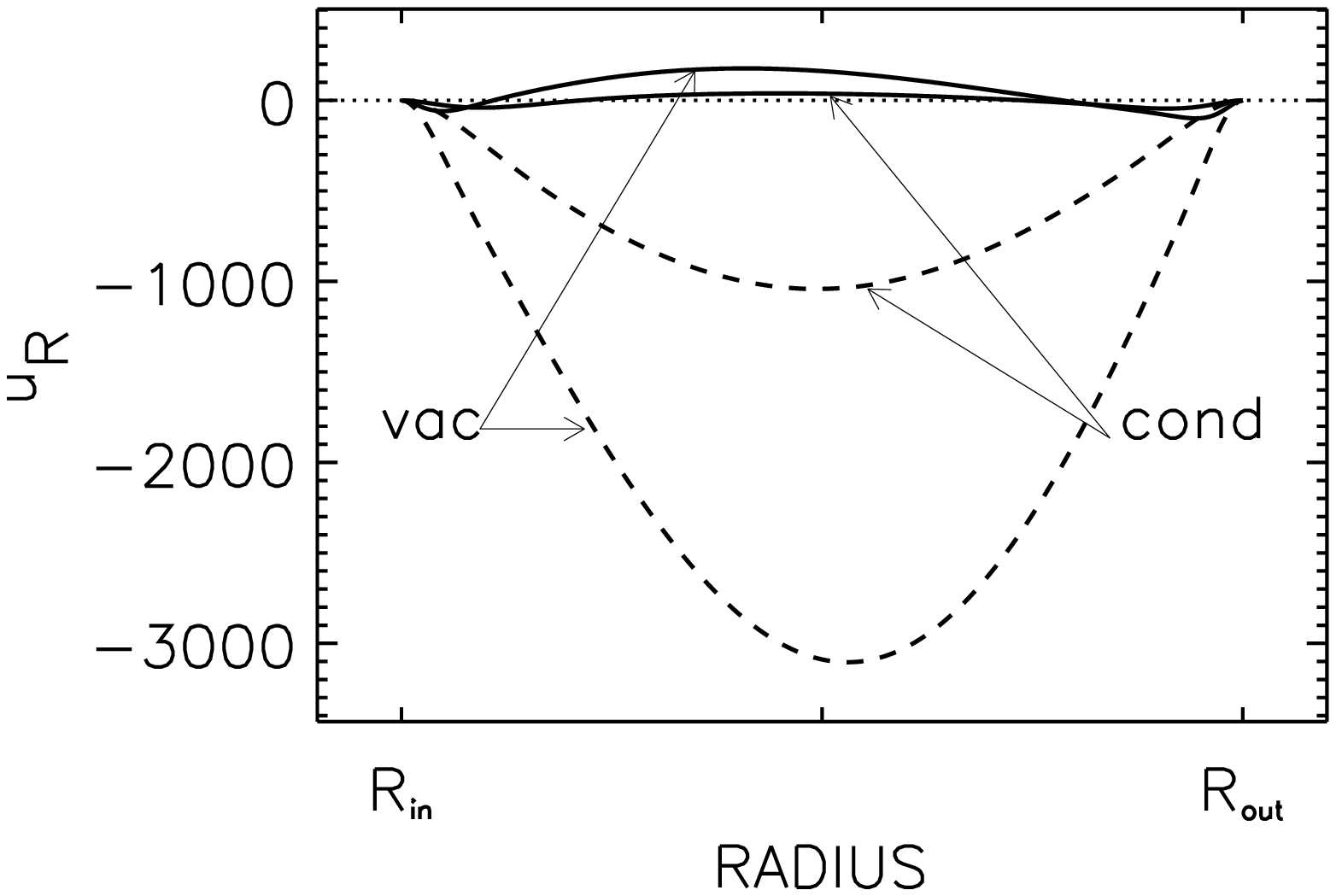} 
 \includegraphics[width=0.33\textwidth]{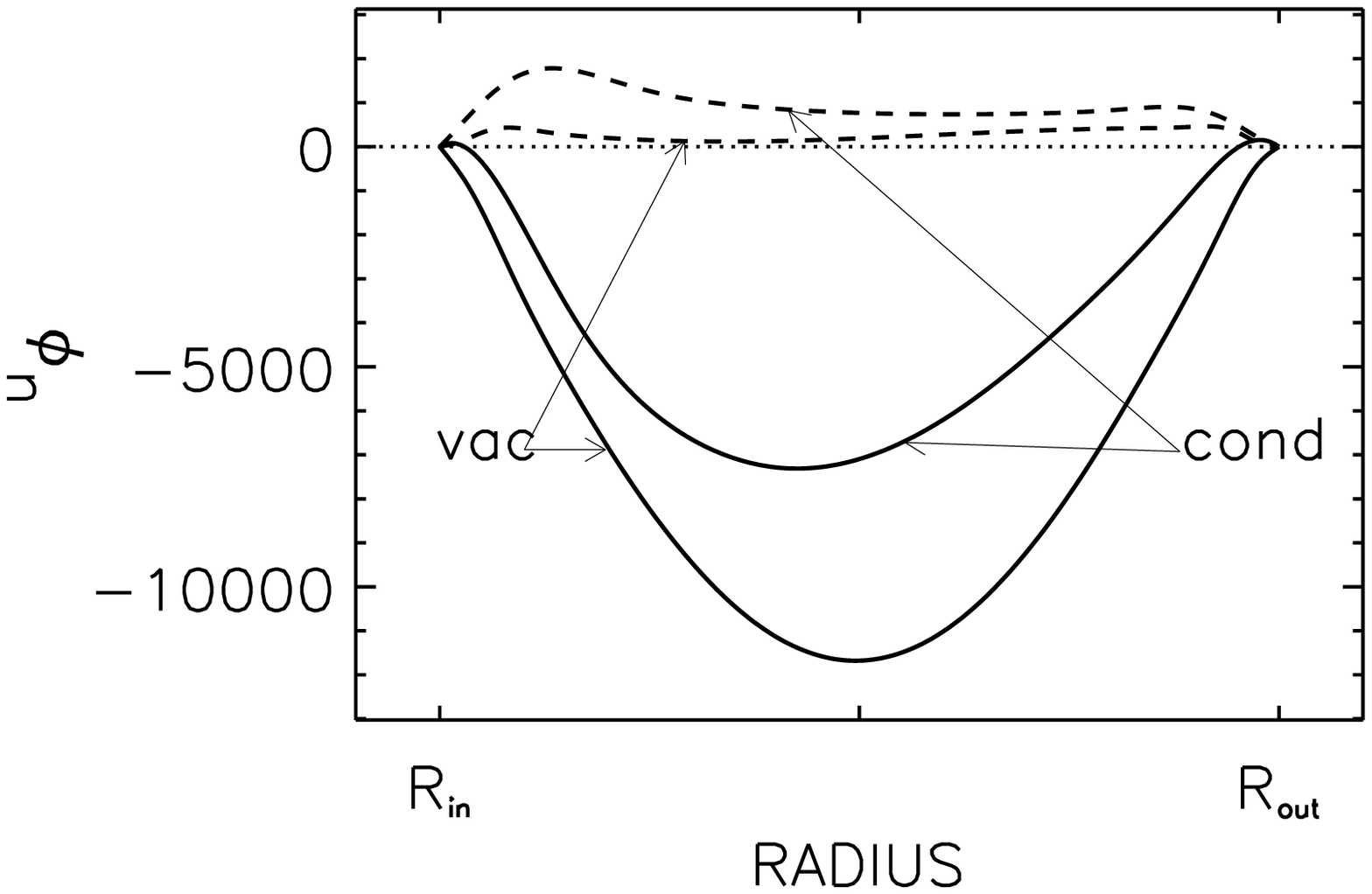} 
 \includegraphics[width=0.33\textwidth]{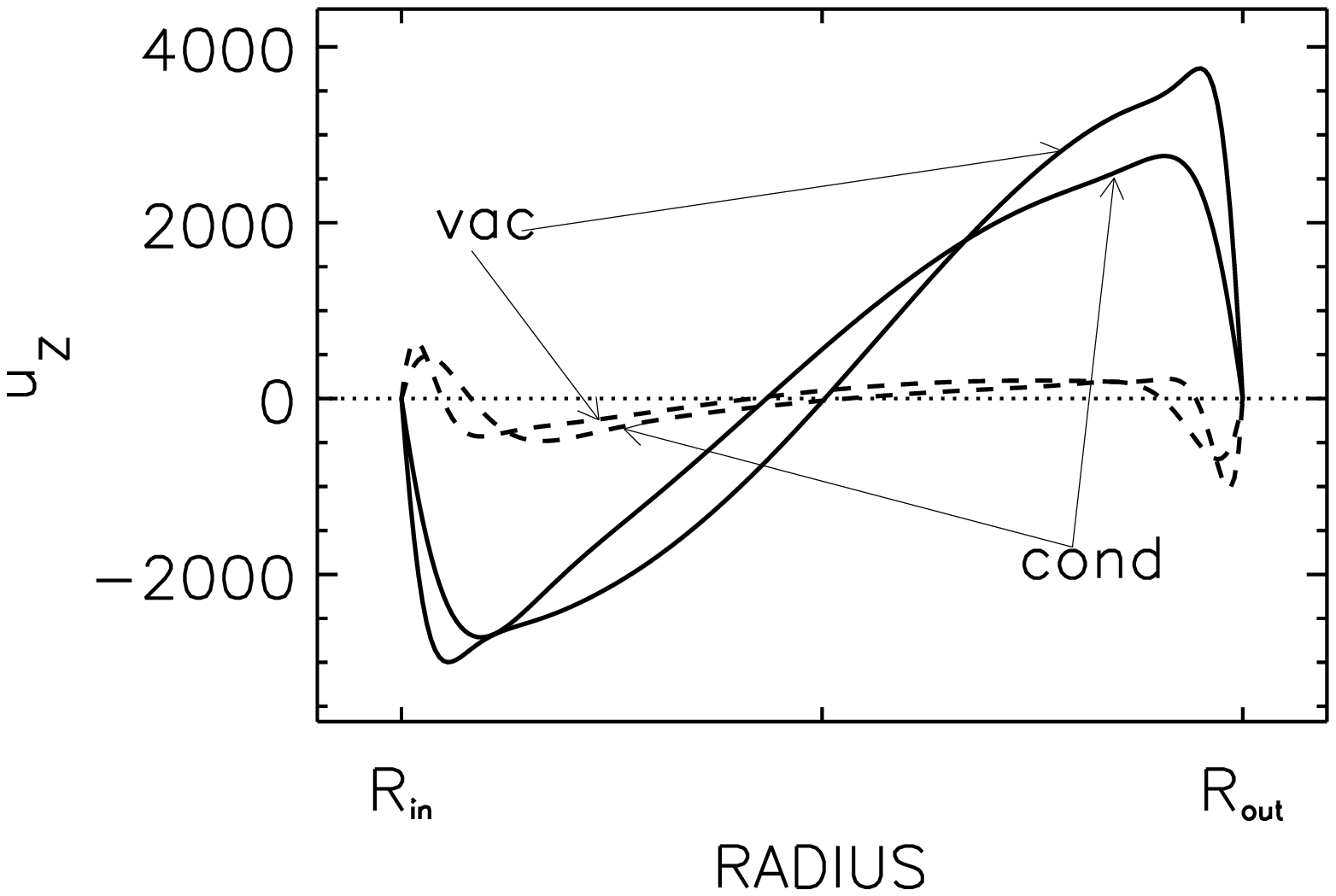} 
  \caption{The real (solid lines) and imaginary (dashed lines) parts of the normalized eigenfunctions for $u_R(R)$, $u_\phi(R)$ and $u_z(R)$  for a super-rotation flow computed at $\Ha=\Hamin$ taken from figure  \ref{fig21}. Insulating boundaries (`vac'), perfectly conducting boundaries (`cond'). $\rin=0.9$. $m= 1$, $\Pm=10^{-5}$, $\mu_B=\rin$, $\mu=5$. } 
\label{functions5} 
\end{figure}

The hydrodynamic rigid-wall conditions of the flow  are fulfilled by the solutions given in figure  \ref{functions5} for super-rotation with $\mu=5$ evaluated with the eigenvalues for the insulating cylinders and also for perfectly conducting cylinders. The equations have been solved exactly for $\Ha=\Hamin$ and the corresponding Reynolds number taken from figure  \ref{fig21} for $\rin=0.9$. Only at this point in the bifurcation map  a unique solution with neutral stability exists. 

Note the $u_\phi$ as much larger than the $u_R$ and the $u_z$ which are of the same order. The $u_R$ is even phase-shifted from the other two components. Hence, the sign of $u_R u_\phi$ takes a double-wave with  two nodes along the azimuth. Another observation concerns the form of the function of the axial flow $u_z(R)$.  It  possesses zeros close to the center of the gap. Downward or upward spirals in $\phi$ and $z$  with a rather small pitch angle are thus characteristic for the flow pattern. In the inner part of the container it is $u_\phi(R) u_z(R) >0$ and in the outer parts it is $u_\phi(R) u_z(R) < 0$. The pattern is thus formed by spiral waves where in a single cell a negative spiral exists in the outer domain and a positive spiral exists in the inner domain. The resulting cells are then vortex-like and the streamlines  seem to be closed in the ($R/z$) plane. 

As expected the magnetic boundary conditions do only slightly influence the geometry of the hydrodynamic flow system. The radial profiles of $u_z$  do hardly   differ from the typical Taylor vortex solution of hydrodynamics.

The expression $u^{\rm R}_Ru^{\rm R}_\phi+u^{\rm I}_Ru^{\rm I}_\phi$ (with the notation $\vec{u}= \vec{u}^{\rm R}+\ri \vec{u}^{\rm I}) $ represents the sign of the  radial flux of the angular momentum averaged over the azimuth $\phi$.  One finds that the small positive real part   $u^{\rm R}_R$  combines with the large negative real  part  $u^{\rm R}_\phi$ (similar for the imaginary parts) to  an {\em inwards} directed flow of angular momentum which is maximal near the center of the gap. The negative sign of this quantity shows that also for super-rotation the instability-induced angular momentum flux tends to reduce the shear of the flow (mainly in the middle of the container). The nonaxisymmetric mode of the radial flux of the angular momentum runs with $-u^{\rm I}_Ru^{\rm R}_\phi \ \cos 2\phi$  with large negative $u^{\rm I}_R u^{\rm R}_\phi$ (see figure  \ref{functions5}). The nonaxisymmetric mode of the angular momentum flow, therefore,  basically exceeds the axisymmetric mode.

  For all magnetic instabilities which for $\Pm\to 0$ scale with $\Ha$ and $\Rey$  the magnetic contribution to the radial angular momentum flow  for small $\Pm$ is  negligible.




\section{Conclusions}
Experimental realizations of magnetorotational instabilities with liquid metals as the fluid conductor often require strong magnetic fields which may exceed achievable laboratory limits. As an example the onset of the azimuthal magnetorotational instability of a Taylor-Couette flow with stationary inner cylinder at $\rin=0.75$ combined with a magnetic field (which is current-free between the cylinders) requires a minimum axial electric current (inside the inner cylinder) of 26 kA if perfectly conducting boundary conditions are used. For insulating cylinder material, however, this limit is much higher (76~kA). The very  large difference between these two numbers requires the knowledge of the true value (\ref{sigma}) for the real conductivity of the cylinder material compared with the conductivity of the liquid fluid between the cylinders.

A general boundary condition in cylindrical geometry for transition zones with discontinuous electric conductivities has thus been formulated also for nonaxisymmetric perturbations. All three components of the magnetic field and the tangential components of the electric field are required to be continuous across the boundary where the conductivity changes. The two cylinders are assumed as thick and as rigidly rotating with prescribed angular velocities. The resulting two boundary conditions (\ref{bc1inner}) and (\ref{bc2inner}) for the inner cylinder and (\ref{bc1outer}) and (\ref{bc2outer}) for the outer cylinder also contain the conditions for perfectly conducting or for insulating material as limiting cases.

There are two main results of the calculations depending on the numerical value of the magnetic Prandtl number. A characteristic example for small ${\rm Pm}=\rO(10^{-5})$ is represented for the case of the double-diffusive instability of super-rotation. Figure \ref{fig31} demonstrates for the model with a quasi-stationary inner cylinder how finite values of $\hats$ determine the Hartmann and Reynolds numbers for the onset of the instability. The critical (i.e. the minimal) Hartmann numbers for perfectly conducting and for insulating boundaries differ by a factor of three. One finds $\hats\simeq 1$ playing a watershed role. For $\hats>1$ the values of $\Hamin$ rapidly approach the lower $\Hamin$ values for perfectly conducting cylinders while they also rapidly approach the larger values for insulating cylinders if $\hats<1$. Note that in the latter case the corresponding Reynolds numbers are smaller than for perfectly conducting cylinders by a factor of about 1.5. The dotted vertical lines in figure  \ref{fig31} demonstrate that there are only small differences of the characteristic minimal Hartmann number and the corresponding Reynolds number of $\hats=5$ (copper cylinders and sodium flow) compared with $\hats\to \infty$.

The second result concerns the quasi-uniform flow, whose    Lundquist numbers $\Lu$ and  magnetic Reynolds numbers $\Rm$ converge  for $\Pm\to 0$, which also means that this flow is {\em stable} in the diffusionless approximation $\Pm=0$. Figures \ref{fig1} and \ref{fig3} represent the unstable solutions for $0.1\leq \Pm\leq 10$. The differences of the extrema for $\hats=0$ and $\hats\to \infty$ remain small despite the large magnetic Reynolds numbers of order $\rO(100)$. It is the magnetic  Prandtl number, however, which defines the extremal values of the Hartmann or the Reynolds numbers as a function of the boundary conditions. For large $\Pm$ a container made from insulating material destabilizes the flow more strongly than a container made from perfectly conducting material. The situation is opposite for $\Pm<1$, where the instability is better supported by perfectly conducting cylinders.

The functions $\Hamin(\hats)$ are always monotonic. At their origin ($\hats=0$) they start with positive slope for large $\Pm$ and with negative slope for small $\Pm$. It is thus clear that the magnetic Prandtl number decides whether insulating or perfectly conducting cylinders better support the instability. In all cases, however, the eigenvalues for $\hats=\rO(10)$ are very close to the eigenvalues for $\hats\to \infty$, and the eigenvalues for $\hats=\rO(0.1)$ are very close to the eigenvalues for $\hats=0$.

\bibliographystyle{gGAF}
\markboth{G.R\"udiger et al.}{Geophysical and Astrophysical Fluid Dynamics}
\bibliography{roberts}
\markboth{G.R\"udiger et al.}{Geophysical and Astrophysical Fluid Dynamics}
\end{document}